\documentclass[sn-mathphys-num]{sn-jnl}

\usepackage{bm,amsmath,empheq,amssymb,graphicx,stmaryrd,float,subcaption,upgreek,sidecap,dsfont,mathtools,gensymb,mathrsfs,xcolor,soul}
\usepackage[abs]{overpic}
\usepackage{hyperref,bookmark}
\usepackage{url}
\usepackage{comment,footnote}
\usepackage[utf8]{inputenc}
\usepackage[T1]{fontenc}
\usepackage{nomencl}
\usepackage{cancel}

\newcommand{\br}{{\bm r}}

\newcommand{\bx}{{\bf x}}
\newcommand{\bff}{{\bf f}}
\newcommand{\bg}{{\bf g}}
\newcommand{\be}{{\bm e}}
\newcommand{\by}{{\bf y}}
\newcommand{\bd}{{\bm d}}
\newcommand{\bu}{{\bm u}}
\newcommand{\bz}{{\bm z}}

\newcommand{\bn}{{\bm n}}
\newcommand{\bmo}{{\bm m}}
\newcommand{\bp}{{\bf p}}

\newcommand{\bP}{{\bm P}}

\newcommand{\bq}{{\bf q}}

\newcommand{\E}{{\mathcal{L}}}

\newcommand{\mcL}{{\mathcal{A}}}

\newcommand{\bzero}{{\bm 0}}

\newcommand{\alert}[1]{{\color{black} #1}}

\newcommand{\eframe}{\{\be_1,\be_2,\be_3\}}
\newcommand{\dframe}{\{\bd_1,\bd_2,\bd_3\}}
\newcommand{\dsframe}{\{\bd_1^\ast,\bd_2^\ast,\bd_3^\ast\}}
\newcommand{\nem}{\bm{q}}
\newcommand{\x}{\bm{x}}
\newcommand{\y}{\bm{y}}
\newcommand{\Lc}{\mathbf{L}}
\newcommand{\Lr}{\mathbf{L}_0}
\newcommand{\C}{\mathbf{C}}
\newcommand{\F}{\mathbf{F}}
\newcommand{\I}{\mathbf{I}}
\newcommand{\trans}{^\mathsf{T}}
\newcommand{\ave}[1]{{\left\langle #1 \right\rangle}}
\newcommand{\sphere}{\mathbb{S}^2}
\newcommand{\surface}{\mathscr{S}}

\DeclareMathOperator{\tr}{tr}
\DeclareMathOperator{\daw}{daw}

\begin{document}
	\title[Work and Activation]{Work and Activation in a Nematic Polymer Network Ribbon}
	

	\author*[1]{\fnm{Harmeet} \sur{Singh}}\email{harmeet.singh@iitgn.ac.in}
	\author[2]{\fnm{Krishnan} \sur{Suryanarayanan}}\email{krishnan.s@iitgn.ac.in}
	\author[3]{\fnm{Epifanio G.} \sur{Virga}}\email{eg.virga@unipv.it}
	\equalcont{These authors contributed equally to this work.}
	
	\affil[1,2]{\orgdiv{Department of Mechanical Engineering}, \orgname{Indian Institute of Technology Gandhinagar}, \orgaddress{\street{}\city{Palaj, Gandhinagar}, \postcode{382055}, \country{India}}}
	\affil*[3]{\orgdiv{Dipartimento di Matematica}, \orgname{Universit\`a di Pavia}, \orgaddress{\street{via Ferrata 5}, \city{Pavia}, \postcode{I-27100}, \country{Italy}}}

\abstract{
We study spontaneous deformations of a ribbon made of nematic polymer networks and activated under the action of a mechanical load.
We show that when such ribbons are activated appropriately, the deformations produced can pull back and perform work against the externally applied load.
We perform two numerical experiments to demonstrate this effect: (1) the \emph{pulling} experiment, where the ribbon is pulled longitudinally by a point force, and (2) the \emph{bending} experiment, where the ribbon is bent out of plane by a terminally applied point force.
We quantify the capacity of the ribbon to work against external loads, and compute its dependence on both the ribbon thickness and the imprinted nematic texture (that is, the distribution of the nematic directors across the ribbon's length).
Finally, we compute the efficiency of the activation process. Building on the outcomes of our numerical explorations, we formulate two educated conjectures on how the activation efficiency can in general be improved by acting on both the applied load and the imprinted nematic texture.
}




\maketitle

\section{Introduction}\label{sec:introduction}
Liquid crystal elastomers (LCEs) are rubber-like materials impregnated with nematogenic molecules, which as parts of the polymeric network composing the material have the potential of affecting the network's shape in consequence of the long-range order that characterizes their molecular organization.

A thorough introduction to these fascinating materials and the theories that describe their behaviour is the already classic book \cite{warner:liquid}; several reviews are also available \cite{mahimwalla:azobenzene,ube:photomobile,white:photomechanical,ula:liquid,pang:photodeformable,kuenstler:light,warner:topographic};  a recent addition to these is \cite{lagerwall:liquid}, which in its witty style and historical perspective makes a very enjoyable reading.

General continuum theories of these materials are available: we only cite \cite{anderson:continuum,zhang:continuum,bai:photomechanical,mihai:nematic}, being fully aware that this is an incomplete list. Applications are likewise manifold; a journal's special issue \cite{korley:introduction} collects a few.

As acutely remarked in \cite{lagerwall:liquid}, LCEs are indeed expected to find much more uses than they presently do: they should only be ``brought out of research labs.'' This is because LCEs are more versatile materials than classic shape memory polymers, which have already many industrial applications \cite{hager:shape,wang:advances}. Unlike the latter, which can change shape only once, LCEs can switch \emph{reversibly} between two programmable shapes under the action of an appropriate external stimulus.

We may say that conventional rubber is a liquid impeded to flow by the entanglement of cross-links that confers to the material a solid-like response. With a parallel metaphor, we may also say that a LCE is liquid crystal suffering a similar fate. The additional degrees of freedom that arise in LCEs compared to conventional elastomers are described by the nematic \emph{director} field $\nem$,\footnote{The symbol $\bn$ is most often used to denote the nematic director; here we deviate from tradition to avoid clash with the typical notation for the internal force in a framed curve, which will be introduced in the following section.} which represents the orientation of liquid-crystalline molecules in their condensed phase, and by a scalar \emph{order parameter} $S$, which represents the degree of orientational order present locally. These extra fields are responsible for the fascinating mechanical behaviour of LCEs, mostly stemming from their coupling  with conventional kinematic measures.

There are several classes of LCEs; we shall be confined here to \emph{nematic} LCEs (actually, to a subclass of these), in which molecules tend to orient alike, but their positions in space are disordered. As remarked in \cite{white:photomechanical}, nematic LCEs are known under a variety of names, not all synonyms. Nominalistic nuances apart, what effectively makes the difference between them (and should be reflected in their appellation) is the extent of cross-linking: the higher this is, the stiffer the material becomes and more is the nematic director $\nem$ linked to the polymer network motion.

When $\nem$ is completely ensalved to the macroscopic deformation, which is the case for extreme cross-linking, LCEs are called \emph{nematic polymer networks} (NPNs).\footnote{They are also called nematic \emph{glasses} in \cite{modes:disclination} and lately also \emph{glassy} liquid crystal networks \cite{lagerwall:liquid}, a newly added name to the already abundant taxonomy of LCEs.}

In this paper, we shall only be concerned with NPNs. Several mechanisms are capable of inducing a change in the degree of order among the nematogenic constituents of the polymer chains in NPNs. They range from heat exchanges modifying the temperature to illumination with frequencies triggering the morphing of dispersed dyes affecting the ordering of the nematic molecules. Light can be  an actuator as good as heat, with the advantage of being able to act contactlessly.\footnote{To describe the interplay between nematic ordering and photon absorption, models with a statistical mechanics twist have also been proposed \cite{bai:photomechanical,corbett:nonlinear,corbett:linear,corbett:polarization,korner:nonlinear,goriely:rod,sonnet:model} (see also \cite{cedeno_madera:understanding} for a recent review).}

In this paper, following the theory put forward in \cite{singh:ribbon}, we shall remain agnostic about the specific mechanism at play to change the ordering in the nematic molecular organization. We shall assume that a uniform scalar order parameter $S_0\ge0$ characterizes the nematic order in the reference configuration and another scalar order parameter $S\ge0$ (still uniform) characterizes the nematic order in the present configuration. Letting $S\neq S_0$ amounts to activating the material. If $S>S_0$, nematic molecules become more aligned along the director $\nem$; if $S<S_0$, they become less oriented along $\nem$. We shall also assume that the cross-linking of the material takes place in the reference configuration, so that $S_0$ and $\nem_0$ are prescribed there. By changing $S_0$ into $S$, the system is brought out of equilibrium and a spontaneous deformation ensues that conveys $\nem_0$ into $\nem$ in a material fashion, so that
\begin{equation}
	\label{eq:deformation_coupling}
	\nem=\frac{\mathbf{F}\nem_0}{|\mathbf{F}\nem_0|}\,,
\end{equation}
where $\mathbf{F}$ is the deformation gradient. $S$ is here  the only activation parameter.

Our setting will \emph{not} be three-dimensional. We shall adopt a theory for thin  NPN sheets that was obtained in \cite{ozenda:blend} by dimension reduction of the celebrated  trace-formula for the elastic free-energy density (per unit volume), which has a long history \cite{blandon:deformation,warner:theory,warner:elasticity,warner:nematic} (see also Chap.\,3 of \cite{warner:liquid}).

The theory presented in \cite{ozenda:blend}, building upon an extension of the classical Kirchhoff-Love hypothesis of plate theory \cite{ozenda:kirchhoff}, arrives at an elastic free-energy density (per unit area) for a thin  NPN sheet featuring two separate \emph{contents}, a \emph{stretching} content scaling like the sheet's \alert{half} thickness $h$, and a \emph{bending} content scaling like $h^3$. This theory will be adopted here in the limit where the NPN sheet under consideration can be regarded as a \emph{narrow ribbon}, as in the previous study \cite{singh:ribbon}. Moreover, as in \cite{singh:bending}, special attention will be devoted to the role played by $h$ in driving stretching and bending deformation modes. The major novelty of the present study lies perhaps in the presence of external mechanical loads that can interfere with the spontaneous deformation prompted by activation.

A NPN ribbon will be described as a \emph{framed} material curve, albeit with a rather unorthodoxal energy. In Sect.~\ref{sec:framed_material_curves}, we find it convenient to rephrase afresh in our context the Lagrangian theory of framed material curves, which will then be used in Sect.~\ref{sec:ribbon} to obtain the system of equilibrium equations for a NPN ribbon, after having recalled in Sect.~\ref{sec:energetics} the energetics of the parent mechanical theory for nematic elastomers in three space dimensions, to make our development self-contained. The governing equations turn out to be so intricate not to be amenable to an analytic treatment; even numerics must be adapted in Sect.~\ref{sec:numerics} to tackle the hybrid character (both differential and finite) of these equations. Section~\ref{sec:a_cantilever_ribbon} is devoted to two numerical experiments where a cantilever ribbon  is pulled or bent prior to activation. Both qualitative and quantitative aspects of the calculated solutions are presented, some being unexpected. In Sect.~\ref{sec:framed_material_curves}, much in the spirit of \cite{zhou:theoretical}, we introduce and compute a measure of actuation efficiency for both experiments, paying special attention to the factors, either geometric or mechanical, that can enhance it. Finally, in Sect.~\ref{sec:conclusions}, we collect the conclusions of this study. 

\section{Lagrangian theory of framed material curves}\label{sec:framed_material_curves}
By a framed material curve we mean a smooth curve $\br(s)\in\mathbb{R}^3$ endowed with an orthonormal frame $\dframe$ of directors $\bd_i(s)$, 
where $s\in[0,L]$ is any scalar parameter whose values identify material points on $\br(s)$.
We restrict our interest to curves where $\bd_3(s)$ is constrained to lie along the tangent $\br'(s)$, with prime being derivation with respect to $s$.
We will refer to such a curve as an \emph{adapted framed curve}, which is completely determined by the set $\mathcal{Q}(s):=\{\br(s),\bd_i(s)\}$ (see Chapt. VIII of \cite{antman:nonlinear}). 
The kinematics of an adapted framed curve are described by the following two relations,
\begin{align}
	\br' = v_3 \bd_3\, ,\qquad \bd'_i = \bu\times\bd_i\, .\label{eq:kinematics}
\end{align}
The first constraint in \eqref{eq:kinematics} ensures that $\bd_3$ is parallel to the tangent, with $v_3(s)>0$, while the second relation ensures the orthonormality of the director frame as it evolves along the paramater $s$.
Vector $\bu$ in \eqref{eq:kinematics}$_2$ is the Darboux vector associated with the frame $\dframe$. 

Consider the following \emph{action} functional $\mcL$ defined on an adapted framed curve,
\begin{align}
	\mcL[\mathcal{Q}(s),f_i,v_3]:= \int_0^L\E(u_i,f_j,u'_i,f'_j,v_3)\, ds + \int_0^L \bn(s)\cdot(\br'-v_3\bd_3)\, ds - \bP\cdot\br(L)\, . \label{eq:Lagrangian}
\end{align}
Here, the \emph{Lagrangian} $\E$ is the energy density function\footnote{We consider $\E$ to depend on the first derivatives of $u_i$ and $f_j$ only, since it fits the description of our functional of interest. However, the method outlined below can be applied to functionals containing higher order derivatives.} associated with the framed curve, $u_i:=\bu\cdot\bd_i$, $i\in\{1,2,3\}$, and $f_j$, $j\in I\subset\mathbb{N}$, are a number  of  additional ancillary functions (indexed in $I$), whose role will be highlighted in the following.
The second integral enforces constraint \eqref{eq:kinematics}$_1$ in a pointwise manner through a Lagrange multiplier function $\bn(s)$.
The third integral accounts for the potential associated with any point force $\bP$ possibly acting at one of the two terminal ends.
Without any loss of generality, we assume that this point force is acting at $s=L$. We shall consider $s$ as a surrogate for \emph{time}; in line with this, we shall often refer to equations involving derivatives in $s$ as \emph{evolution} equations.

Our objective is to minimize the action functional $\mcL$ in  \eqref{eq:Lagrangian} in the set of  all \emph{admissible} configurations consistent with the boundary conditions at hand.
While the procedure we outline does not depend on the nature of the prescribed boundary conditions, we will assume the following boundary conditions which arise in the specific problems we wish to study below,
\begin{align}
	\mathcal{Q}(0)=\{\br_0,\be_i\},\label{eq:boundary_condition}
\end{align}
where $\br_0$ is an assigned  vector and $\eframe$ is a given Cartesian frame.

Consider the following variations in the configuration of an adapted framed curve,
\begin{align}
	\br^*:=\br+\delta\br\,,\quad\bd^*_i:=\bd_i + \delta\bd_i\,,\quad f^*_j:= f_j+\delta f_j\,,\quad v^*_3:=  v_3+\delta v_3\, ,\label{eq:variations}
\end{align}
where $\delta\br$ is subject to the condition $\delta\br(0)=0$ due to \eqref{eq:boundary_condition}, whereas $\delta f_j$ and $\delta v_3$ are arbitrary.
The variations $\delta\bd_i$ must be such that the frame $\dsframe$ is also orthonormal.
This constrains the variations $\delta\bd_i$ to be of the form,
\begin{align}
	\delta\bd_i = \delta\bz \times \bd_i\, ,\label{eq:director_variation_constraint}
\end{align}
where $\delta\bz$ is an abitrary vector-valued function.
The variation $\delta\bd_i$ will further induce a variation in the Darboux vector,
\begin{align}
	\bu^*:=\bu + \delta\bu\, .\label{eq:Darboux_vector_variation}
\end{align}
To express $\delta\bu$ in terms of $\delta\bz$, with the aid of \eqref{eq:director_variation_constraint} and \eqref{eq:kinematics}$_2$ we transform the equation $\left(\bd_i^*\right)' = \bu^*\times\bd_i^*$ into the identity
\begin{align}\label{eq:pre_identity}
	\left(\delta\bz' - \bu\times\delta\bz - \delta\bu\right)\times\bd_i = {\bm 0}\, .
\end{align}
Since $\dframe$ is a basis, we conclude from \eqref{eq:pre_identity} that
\begin{align}
	\delta\bu = 	\delta\bz' - \bu\times\delta\bz\, .\label{eq:variation_Darboux_vector}
\end{align}
Using \eqref{eq:variation_Darboux_vector} and \eqref{eq:director_variation_constraint}, the variations $\delta u_i:= \delta(\bu\cdot\bd_i)$ can be written as
\begin{align}
	\delta u_i&=\delta\bz'\cdot\bd_i\, .\label{eq:variation_Darboux_component}
\end{align}

Next, the variation of the action functional \eqref{eq:Lagrangian}, induced by variations \eqref{eq:variations}, can be computed in the standard way as,
\begin{align}
	\delta\mcL &= \int_0^L\left[\frac{\partial\E}{\partial u_i}\delta u_i + \frac{\partial\E}{\partial f_j}\delta f_j + \frac{\partial\E}{\partial u'_i}\delta u'_i + \frac{\partial\E}{\partial f'_j}\delta f'_j + \frac{\partial\E}{\partial v_3}\delta v_3\right] ds\nonumber\\ &+ \int_0^L\left[\bn\cdot\left(\delta\br' - \delta v_3\bd_3 - v_3\delta\bd_3\right)\right]ds - \bP\cdot\delta\br(L)\, ,\label{eq:variation_Lagrangian}
\end{align}
where sum over repeated indices is understood.
On integrating by parts, and using \eqref{eq:director_variation_constraint} and \eqref{eq:variation_Darboux_component}, $\delta\mcL$ can be written as,
\begin{gather}
	\delta\mcL = \int_0^L\bigg\{-\bn'\cdot\delta\br- \left[\left(\left[\frac{\partial\E}{\partial u_i} - \left(\frac{\partial\E}{\partial u'_i}\right)'\right]\bd_i\right)' + \left(v_3\bd_3\times\bn\right)\right]\cdot\delta\bz\nonumber\\ +\left[\frac{\partial\E}{\partial f_j}\right. 
	\left.- \left(\frac{\partial\E}{\partial f'_j}\right)'\right]\delta f_j + \left[\frac{\partial\E}{\partial v_3}-\bn\cdot\bd_3\right]\delta v_3\bigg\} ds\nonumber\\
	+\left(\bn\cdot\delta\br\right)\big|_{s=0}^{s=L} - \bP\cdot\delta\br(L)+ \left(\frac{\partial\E}{\partial u'_i}\bd_i\cdot\delta \bz'\right)\bigg|_{s=0}^{s=L}\nonumber\\
	+ \left(\left[\frac{\partial\E}{\partial u_i} - \left(\frac{\partial\E}{\partial u'_i}\right)'\right]\bd_i\cdot\delta\bz\right)\bigg|_{s=0}^{s=L}+ \left(\frac{\partial\E}{\partial f'_j}\delta f_j\right)\bigg|_{s=0}^{s=L}\, .\label{eq:variation_Lagrangian_final}
\end{gather}
To require that $\delta\mcL=0$, we invoke the arbitrariness of the variations in the bulk and we write down the governing equations in the following transparent form,
\begin{subequations}\label{eq:balance_laws}
	\begin{align}
		\bn' &= \bzero\, ,\label{eq:force_balance}\\
		\bmo' + \br'\times\bn &= \bzero\, ,\label{eq:moment_balance}\\
		\frac{\partial\E}{\partial f_j} - \left(\frac{\partial\E}{\partial f'_j}\right)'&=0\, ,\label{eq:EL_function}\\
		\frac{\partial\E}{\partial v_3}-\bn\cdot\bd_3&=0\, ,\label{eq:EL_stretch}
	\end{align}
\end{subequations}
where,
\begin{align}
	\bmo:=m_i\bd_i\qquad\text{and}\qquad m_i:=\frac{\partial\E}{\partial u_i} - \left(\frac{\partial\E}{\partial u'_i}\right)'\, .\label{eq:constitutive_law}
\end{align}
We have used $\br' = v_3\bd_3$ in writing \eqref{eq:moment_balance}.
As can be clearly seen, equations \eqref{eq:force_balance} and \eqref{eq:moment_balance} are formally the Kirchhoff rod equations of force and moment balance, accompanied by (what can be interpreted as) constitutive relations \eqref{eq:constitutive_law} for the internal moment.
When the energy density $\E$ is independent of $u'_i$, the constitutive relations \eqref{eq:constitutive_law} reduce to the standard constitutive relation for a hyperelastic rod (see again Chap. VIII of \cite{antman:nonlinear}).
Also, one can interpret \eqref{eq:EL_stretch} as a constitutive relation for $n_3:=\bn\cdot\bd_3$.
Equations \eqref{eq:force_balance} and \eqref{eq:moment_balance} are the vector representations of equations (15) and (16) of Hornung's article \cite{hornung:euler}, and equations (32) and (33) of Starostin's article \cite{starostin:theory}.
Equation \eqref{eq:constitutive_law} corresponds to equation (14) of \cite{hornung:euler}, and equation (34) of \cite{starostin:theory}.

To obtain the boundary conditions, we note that \eqref{eq:boundary_condition} and \eqref{eq:director_variation_constraint} imply that $\delta\br(0)=\bm 0$, $\delta\bz(0)=\bm 0$, using which we conclude from \eqref{eq:variation_Lagrangian_final} that
\begin{subequations}\label{eq:boundary_conditions}
	\begin{align}
		\bn(L)&=\bP\, ,\label{eq:boundary_conditions_a}\\
		\bmo(L)&={\bm 0}\, ,\label{eq:boundary_conditions_b}\\
		\frac{\partial\E}{\partial u'_i}\bd_i\bigg|_{s=0} = \frac{\partial\E}{\partial u'_i}\bd_i\bigg|_{s=L} &={\bm 0}\, ,\label{eq:boundary_conditions_c}\\
		\frac{\partial\E}{\partial f_j}\bigg|_{s=0}=\frac{\partial\E}{\partial f_j}\bigg|_{s=L}&=0\, .\label{eq:boundary_conditions_d}
	\end{align}
\end{subequations}
Equations \eqref{eq:balance_laws}-\eqref{eq:boundary_conditions} form the basis of our general Lagrangian theory for framed material curves, where time is replaced by arc-length. In Sect.~\ref{sec:ribbon}, we shall apply it to the deformation of an activable soft material, after a brief digression about the energetics of nematic elastomers.

\section{Energetics of nematic elastomers}\label{sec:energetics}
Our theory is based on the \emph{trace formula} for the elastic free-energy density $f_\mathrm{e}$ (per unit volume of the reference configuration) proposed in \cite{finkelmann:elastic} for nematic elastomers,
\begin{equation}
	\label{eq:free_energy_density_elastomers}
	f_\mathrm{e}=\frac12 n_\mathrm{s}k_\mathrm{B}T_0\bigg\{\tr(\F\trans\Lc^{-1}\F\Lr)+\ln\left(\frac{\det\Lc}{\det\Lr}\right)\bigg\}\,,
\end{equation}
where $n_\mathrm{s}$ is the number of polymer strands per unit volume, $T_0$ is the temperature in the reference configuration, $\F$ is the deformation gradient, and $\Lc$ and $\Lr$ are the \emph{step-length} tensors in the present and reference configurations, respectively.

The latter tensors reflect the organization of nematogenic molecules within the polymer network. Following \cite{verwey:elastic,nguyen:theory}, we represent them as
\begin{equation}
	\label{eq:L_tensors}
	\Lr=A_0(\I+S_0\nem_0\otimes\nem_0)\quad\text{and}\quad\Lc=A(\I+S\nem\otimes\nem)\,,
\end{equation}
where $\I$ is the identity tensor (in three-dimensional space), $\nem_0$ and $\nem$
are the nematic directors in the reference and present configurations, and $S_0$ and $S$ the corresponding scalar order parameters. These parameters are related to the nematic scalar order parameters $Q_0$ and $Q$ of the classical Maier-Saupe theory through the equations,
\begin{equation}
	\label{eq:S_and_Q}
	S_0=\frac{3Q_0}{1-Q_0}\quad\text{and}\quad S=\frac{3Q}{1-Q}\,,
\end{equation}
while the amplitudes $A_0$ and $A$ are given by
\begin{equation}
	\label{eq:A_and_Q}
	A_0=l(1-Q_0)\quad\text{and}\quad A=l(1-Q)\,,
\end{equation}	
where $l$ is the length of the rod representing a single monomer in the classical statistical mechanics model that describes a polymer strand as a chain of freely jointed equal rods. It follows from \eqref{eq:S_and_Q} and \eqref{eq:A_and_Q} that 
\begin{equation}
	\label{eq:Q_and_S}
	Q_0=\frac{S_0}{3+S_0}\,,\quad Q=\frac{S}{3+S}\,,\quad\frac{A}{A_0}=\frac{3+S_0}{3+S}\,.
\end{equation}

The nematic scalar order parameter $Q$ is defined as
\begin{equation}
	\label{eq:Q_definition}
	Q:=\ave{P_2(\nem\cdot\bm{\ell})}\,,
\end{equation}
where $P_2(x)$ is the second Legendre polynomial, $\bm{\ell}\in\sphere$ is the unit vector designating the orientation of a single monomer, and the brackets $\ave{\cdots}$ denote ensemble average. The same formula applies to $Q_0$, with $\nem$ replaced by $\nem_0$. It is an easy consequence of \eqref{eq:Q_definition} that $Q$ ranges in the interval $[-\frac12,1]$, the upper end designating the (ideal) state of perfect orientation of all molecules along $\nem$ and lower end representing the state of isotropic orientation in the plane orthogonal to $\nem$. Correspondingly, by \eqref{eq:S_and_Q}, $S\in[-1,\infty[$. Here both $S_0$ and $S$ will be taken as positive.

Changing the degree of order (measured by either $S$ or $Q$) costs energy, which is accounted for by the condensation free-energy density (per unit volume) $f_\mathrm{c}$ arrived at in the mean-field theory of Maier and Saupe~\cite{maier:einfache,maier:simple}, as described, for example, in Sect.~1.3 of \cite{sonnet:dissipative},
\begin{equation}
	\label{eq:condensation_energy}
	f_\mathrm{c}=n_\mathrm{n}k_\mathrm{B}T_0U_\mathrm{MS}(\beta;Q),
\end{equation}
where $n_\mathrm{n}$ is the number density of nematogenic molecules and 
\begin{equation}\label{eq:U_MS}
	U_\mathrm{MS}(\beta;Q):=\beta\left(\frac13Q^2-\frac23Q\right)-\ln\left(\frac{\daw(\sqrt{\beta Q})}{\sqrt{\beta Q}}\right)\,.
\end{equation}
In \eqref{eq:U_MS}, $\beta$ is the Maier-Saupe \emph{molecular interaction} energy (scaled to $k_\mathrm{B}T_0$) and $\daw$ is the Dawson integral, defined as
\begin{equation}
	\label{eq:dawson}
	\daw(x):=e^{-x^2}\int_0^xe^{t^2}dt\quad\text{for}\ x\in\mathbb{R}.
\end{equation}
Moreover, $Q$ is related to $S$ as in \eqref{eq:Q_and_S}.

The absolute minimizer of $U_\mathrm{MS}$ depends on $\beta$; for $\beta=\beta_0$, it is equal to $Q_0$, as delivered by \eqref{eq:Q_and_S} in terms of the scalar order parameters $S_0$ imprinted in the reference configuration at the time of cross-linking. 

Here, for simplicity, we shall consider $S$ as a control parameter, ascribing to external causes the change in the value of $\beta$ away from $\beta_0$ which is responsible for shifting the absolute minimizer of the potential $f_\mathrm{c}$ from $Q_0$ to $Q$.

Since also $f_\mathrm{e}$ depends on $S$, a more orthodox approach would be to use $\beta$ as a control parameter and let the total free-energy density $f_\mathrm{t}:=f_\mathrm{e}+f_\mathrm{c}$ decide the equilibrium value of $S$ in competition with the elastic deformation. Our approach is however justified by the assumption that the elastic deformation is a much slower response compared to the activation processes that we envision, which will be taken as virtually instantaneous. Under this assumption, were we only interested in the mechanical response of the system, we could safely disregard all energies that depend only on $S$, as done for example in \cite{singh:ribbon,singh:bending}. Here, for use in Sect.~\ref{sec:efficiency}, we keep all these energies as well, as we are also interested in the estimate of the activation energy and the efficiency of the mechanical yield.

In summary, we write the total free-energy density as
\begin{align}
	\label{eq:free_energy_total}
	f_\mathrm{t}=\frac12n_\mathrm{n}k_\mathrm{B}T_0\bigg\{&\gamma\frac{3+S}{3+S_0}\bigg[\tr\C+\frac{S_0}{1+S}\nem_0\cdot\C\nem_0-\frac{S}{1+S}\frac{\alert{\nem_0}\cdot\C^2\nem_0}{\alert{\nem_0\cdot\C\nem_0}}\bigg]\nonumber\\
	&+3\gamma\ln\left(\frac{3+S_0}{3+S}\right)+\gamma\ln\left(\frac{1+S}{1+S_0}\right)+2U_\mathrm{MS}(\beta(S);Q(S))\bigg\},
\end{align}
where $\C:=\F\trans\F$ is the \alert{(three-dimensional)} right Cauchy-Green tensor and use has also been made of the kinematic constraint \eqref{eq:deformation_coupling} (see also \cite{singh:ribbon,singh:bending}). In \eqref{eq:free_energy_total}, $Q(S)$ is the function in \eqref{eq:Q_and_S} and $\beta(S)$ is such that $U_\mathrm{MS}$ is minimized by $Q(S)$. The parameter 
\begin{equation}
	\label{eq:gamma}
	\gamma:=\frac{n_\mathrm{s}}{n_\mathrm{n}}
\end{equation}
describes the degree of cross-linking in the material: the larger $\gamma$, the stronger the cross-linking. Following \cite{corbett:polarization}, for strongly cross-linked polymers, such as NPNs, we shall choose $\gamma=1/10$.\footnote{\alert{As remarked in \cite{white:photomechanical}, there is sufficient experimental evidence \cite{ware:programmable} to hold that the traditional distinction between nematic elastomers and nematic polymer networks is obsolete: the mechanical response of these materials is a continuum dictated by the extent of cross-linking. According to \cite{corbett:polarization}, a more weakly cross-linked material would have $\gamma=1/50$.}}

\alert{A method of dimension reduction developed in \cite{ozenda:blend} showed how to convert the core of the trace formula,
\begin{equation}
	\label{eq:core_energy}
	F(\C):=\tr(\F\trans\Lc^{-1}\F\Lr)\,,
\end{equation}
into a surface energy when the material is confined to a sheet of thickness $2h$ around a midsurface $\surface_0$ in the $(x_1,x_2)$ plane of the reference space. It was proved that 
\begin{equation}
	\label{eq:dimension_reduction}
	\int_{-h}^{+h}F(\C)dx_3=hf_1+h^3f_3+O(h^5)\,,
\end{equation}
where
\begin{subequations}\label{eq:f_1_f_3}
\begin{align}
	f_1&:=2\bigg\{1+\frac{1}{1+S}\bigg[\tr\C_2+S_0\nem_0\cdot\C_2\nem_0+\frac{S}{\nem_0\cdot\C_2\nem_0}\bigg]\bigg\}\,,\label{eq:f_1}\\
	f_3&:=\frac23\bigg\{2(8H^2-K)+\frac{1}{1+S}\bigg[\bigg(\frac{3S}{a_0^2}-a_0^2S-\tr\C_2\bigg)\bigg]K-\frac{4S}{a_0^2}(2H-\kappa_q)\kappa_q\bigg\}\,.\label{eq:f_3}
	\end{align}
\end{subequations}
In equations \eqref{eq:f_1_f_3}, $\C_2$ is the two dimensional right Cauchy-Green tensor defined on $\surface_0$, $a_0^2:=\nem_0\cdot\C_2\nem_0$, $H$ and $K$ are the mean and Gaussian curvatures of the deformed configuration $\surface$ of the sheet's midsurface, and $\kappa_q:=\nem\cdot\mathbf{K}\nem$, where $\mathbf{K}$ is the curvature tensor of $\surface$,
see also equations (73) of \cite{ozenda:blend} and (11) of \cite{singh:ribbon}.

Other theories have been proposed for the free-energy of nematic elastomers, such as the one presented in \cite{mihai:nematic}, which mimics the elastic theory of growth in \cite{rodriguez:stress-dependent}. According to this alternative theory for nematic elastomers, $f_\mathrm{e}$ in \eqref{eq:free_energy_density_elastomers} would be replaced by
\begin{equation}
	\label{eq:W_e}
	W_\mathrm{e}(\F,\nem):=W(\mathbf{A})\,,
\end{equation}
where $W$ is the strain energy of the underlying isotropic polymer network (delivered, for example, by the neo-Hookian formula) and
\begin{equation}
	\label{eq:A_definition}
	\mathbf{A}:=\mathbf{G}^{-1}\F\,,
\end{equation}
where 
\begin{equation}
	\label{eq:G_definition}
	\mathbf{G}{:=a^{-1/6}}\I+\left(a^{1/3} -a^{-1/6}\right)\nem\otimes\nem
\end{equation}
is the \emph{spontaneous} deformation tensor, playing a role akin to $\Lc$ in our theory. This theory has also been applied to rods in \cite{goriely:rod}, following a method for dimensional reduction that had been developed in \cite{moulton:morphogenesis} for the theory of growth. There are differences between this theory and ours: one distinctive feature of our Lagrangian, which  does not appear to be shared by the growth-like Lagrangian, is the dependence of the bending energy $f_3$ on both the stretching measures of the present shape $\surface$  and the nematic orientation $\nem$ on it, besides the measures of curvature (which were of course expected).

The goal of this paper is to estimate the mechanical efficiency of the activation process. With this in mind,  in the following section, we shall see how the free-energy densities in \eqref{eq:f_1_f_3} simplify for a ribbon.}

\section{A nematic polymer network ribbon}\label{sec:ribbon}
\begin{figure}[h]
	\centering
	\includegraphics[width=0.9\textwidth]{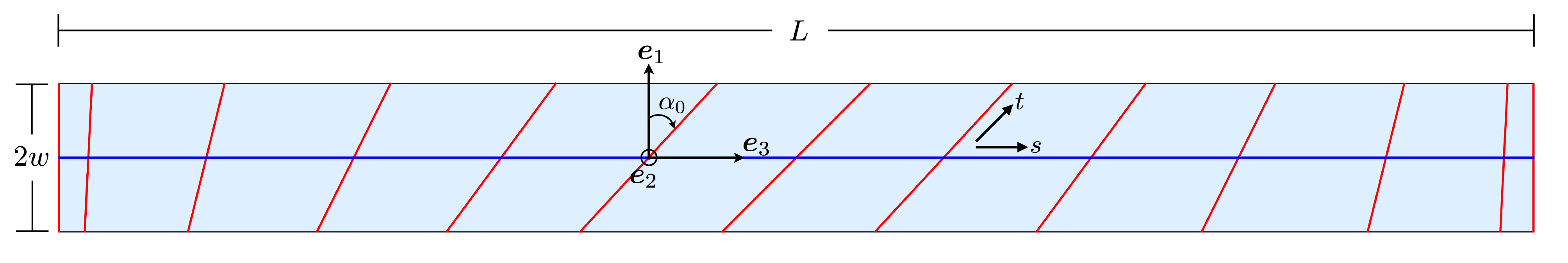}
	\caption{Schematic of the reference configuration of a narrow, rectangular, nematic polymer network ribbon with width $2w$ and length $L$. The coordinate $s$ runs along the centerline, while the coordinate $t$ spans the ribbon's width along the material lines (marked in red) where the nematic director $\nem_0$ has been imprinted prior to activation.}
	\label{fig:generic}
\end{figure}

Here we consider a ribbon comprised of nematic polymer networks. We assume  that it is sufficiently narrow to be treated within our theory for framed curves. The reference configuration, a rectangle of width $2w$ and length $L$, is depicted in Fig.~\ref{fig:generic}.

\alert{As shown in \cite{singh:ribbon} [see, in particular, their equations (43b), (47a), and (51)], for a rectangular ribbon the formula for $f_3$ in \eqref{eq:f_3} simplifies considerably, as both $K$ and $\kappa_q$ vanish. When $\surface_0$ is a rectangular ribbon, by applying the method illustrated in \cite{singh:bending} [see, in particular, their equations (43) and (44)], the total (scaled) free-energy density (per unit normalized length) $\mathcal{F}_\mathrm{t}$ associated with $f_\mathrm{t}$ in \eqref{eq:free_energy_total} can be given the following expression,
}
\begin{gather}
	\mathcal{F}_\mathrm{t}(u_1,\alpha,v_3):=\frac{3+S}{3+S_0}\bigg\{ \frac43\gamma h^3\frac{u_1^2}{v_3^2\cos^4\alpha}+\gamma h\bigg(\alert{1+}\frac{v_3^2\cos^2\alpha}{\cos^2\alpha_0}+\frac{1+S_0}{1+S}\frac{\cos^2\alpha_0}{v_3^2\cos^2\alpha}\nonumber\\+\frac{(v_3^2\cos\alpha\sin\alpha-\cos\alpha_0\sin\alpha_0)^2}{(1+S)v_3^2\cos^2\alpha_0\cos^2\alpha}\bigg)\bigg\}\nonumber\\
	+h\bigg[3\gamma\ln\left(\frac{3+S_0}{3+S}\right)+\gamma\ln\left(\frac{1+S}{1+S_0}\right)+2U_\mathrm{MS}(\beta(S);Q(S))\bigg]\,.\label{eq:original_functional}
\end{gather}
Here $S_0$ and $S$ represent the scalar order parameter in the reference and present configurations, respectively, and $\alpha_0$ and $\alpha$ describe the orientation relative to the ribbon's centerline of the nematic director in the reference and present configurations, respectively.

\alert{It is perhaps worth noting that the bending energy featuring in \eqref{eq:original_functional}, easily recognized as the one scaling like $h^3$, does not only depend on the bending strain $u_1$, but also on the stretching measure $v_3$ and the orientation $\alpha$ of the nematic field in the present configuration.}

In \eqref{eq:original_functional} and in the following, energies are rescaled to the characteristic energy\footnote{The scaling factor $E_0$ of the energy differs slightly from the one, $e_0$, employed in \cite{singh:ribbon}; the two are related by the equation
	\begin{equation*}
		e_0=\frac{3+S}{3+S_0}E_0\,,
	\end{equation*}
	which is recorded for ease of future reference.}
\begin{equation}
	\label{eq:scaling_energy}
	E_0:=2wL^2n_\mathrm{n}k_\mathrm{B}T_0\,.
\end{equation}
Moreover, all lengths are scaled to $L$, including $h$, which represents the ribbon's thickness. Thus, in particular, $s$ will range in $[0,1]$ and $h\ll1$, as dictated by the small-thickness assumption adopted in \cite{ozenda:blend,singh:ribbon}. As customary in plate theory, bending and stretching energies are weighted one relative to the other as $h^3$ versus $h$; all energies of order $h^5$ or higher are neglected.
\alert{The latter is indeed the only approximation made here. We shall not treat stretching and bending energies in a hierarchal manner on account of their different scaling  with $h$, as we are interested in the interplay between these energies as $h$ is varied.}

We regard $S$ as a control parameter that says whether the nematic order has been enhanced ($S>S_0$) or depressed ($S<S_0$) by external agents relative to the order imprinted in the reference configuration. Correspondingly, the angle $\alpha_0$ designates the orientation of the nematic director $\nem_0$ imprinted in the reference configuration (see Fig.~\ref{fig:generic}); it varies with the arc-length $s$ along the centerline of the ribbon and affects the way a given change from $S_0$ to $S$ induced by activation affects the elastic response of the ribbon.

In the present setting of a rectangular ribbon with width $2w$ and length $L$ depicted in Fig.~\ref{fig:generic},
\begin{equation}
	\label{eq:n_0}
	\nem_0(s)=\cos\alpha_0(s)\be_1+\sin\alpha_0(s)\be_3\,,
\end{equation}
while the position vector $\x(s,t)$ is represented by the mapping
\begin{equation}
	\label{eq:x_mapping}
	\x(s,t)=s\be_3+t\nem_0(s)\quad\text{with}\quad-\frac{w}{\cos\alpha_0(s)}\leq t\leq\frac{w}{\cos\alpha_0(s)}.
\end{equation}
The angle $\alpha$ designates the orientation of the nematic director $\nem$ in the present configuration,
\begin{equation}
	\label{eq:n}
	\nem(s)=\cos\alpha(s)\bd_1(s)+\sin\alpha(s)\bd_3(s).
\end{equation}
In a NPN, an area-preserving deformation $\y$, which is defined on a surface and takes values in three-dimensional space \cite{ozenda:blend}, conveys the nematic director as a material line, as prescribed by \eqref{eq:deformation_coupling}. This, combined with the kinematic analysis in \cite{singh:ribbon} and \cite{singh:bending}, leads us to conclude that $\y$ is described by the mapping
\begin{equation}
	\label{eq:y_mapping}
	\y(s,t)=\br(s)+a(s)t\nem(s),
\end{equation}
where
\begin{equation}
	\label{eq:a_definition}
	a(s):=\frac{\cos\alpha_0}{v_3\cos\alpha}.
\end{equation}
It thus remains fully justified that the elastic energy density $\mathcal{F}_\mathrm{t}$ in \eqref{eq:original_functional}, which describes a blend of stretching and bending energy contents, depends on the kinematic variables $(u_1,\alpha,v_3)$ and has $(\alpha_0,S_0,S)$ as parameters.

As shown in \cite{singh:bending} [see, in particular, their equation (46)], the incompressibility constraint is written as,
\begin{align}
	\mathcal{C}(u_2,\alpha,\alpha',v_3):=u_2 - \alpha' + \frac{\alpha'_0 v_3^2 \cos^2\alpha}{\cos^2\alpha_0}=0\, ,\label{eq:original_constraint}
\end{align}
\alert{which relates $u_2$ to other kinematic variables. Similarly, $u_3$ is expressed as
\begin{equation}
	\label{eq:u_3}
	u_3=u_1\tan\alpha\,,
\end{equation}
see also equations (31) of \cite{singh:bending}.
}
In \eqref{eq:original_functional} \eqref{eq:original_constraint}, and  \eqref{eq:u_3}, we perform a change of variables replacing $\alpha$ with  $\eta:=\tan\alpha$.
Consequently, $\alpha_0$ is replaced by $\eta_0:=\tan\alpha_0$.
The resulting functional $\mcL$ in \eqref{eq:Lagrangian}, augmented with the transformed representation of \eqref{eq:original_constraint}, is represented by the Lagrangian
\begin{gather}
	\E(u_1,u_2,\eta,\eta',\mu,v_3) :=\frac{3+S}{3+S_0}\bigg\{ \frac{4}{3}\gamma h^3\frac{u_1^2}{v_3^2}\left(1+\eta^2\right)^2 \nonumber\\+ \gamma h\bigg[\alert{1+}\frac{v_3^2(1+\eta_0^2)}{\left(1+\eta^2\right)} + \frac{1+S_0}{1+S}\frac{\left(1+\eta^2\right)}{v_3^2(1+\eta_0^2)}
		 + \left(v_3^2{\frac{\eta}{(1+\eta^2)}} - \frac{\eta_0}{(1+\eta_0^2)}\right)^2\frac{\left(1+\eta^2\right)(1+\eta_0^2)}{(1+S)v_3^2}\bigg]\bigg\}\nonumber\\
	+h\bigg[3\gamma\ln\left(\frac{3+S_0}{3+S}\right)+\gamma\ln\left(\frac{1+S}{1+S_0}\right)+2U_\mathrm{MS}(\beta(S),Q(S))\bigg]\nonumber\\
	+ \mu(s)\left(u_2 - \frac{\eta' - v_3^2\eta'_0}{1+\eta^2}\right)\, ,\label{eq:combined_functional}
\end{gather}
where $\mu(s)$ is a pointwise Lagrange multiplier function corresponding to the incompressibility constraint. In \eqref{eq:combined_functional}, $\eta$ plays the role of a single ancillary function of the many $f$'s featuring in the general form of $\mcL$ in \eqref{eq:Lagrangian}.

Similarly, the constitutive equations \eqref{eq:constitutive_law} are changed as follows by the intervention of the new variable $\eta$,
\begin{subequations}\label{eq:transformed_constitutive}
	\begin{align}
		m_1 &= \frac{\partial\E}{\partial u_1} + \frac{\partial\eta}{\partial u_1}\left[\frac{\partial\E}{\partial\eta} - \frac{d}{ds}\left(\frac{\partial\E}{\partial\eta'}\right)\right] \label{eq:m1_transformed}\, ,\\
		m_2 &=\frac{\partial\E}{\partial u_2}\label{eq:m2_transformed}\, ,\\
		m_3&=\frac{\partial\eta}{\partial u_3}\left[\frac{\partial\E}{\partial\eta} - \frac{d}{ds}\left(\frac{\partial\E}{\partial\eta'}\right)\right]\, .\label{eq:m3_transformed}
	\end{align}
\end{subequations}

\alert{In the following section, we shall write the equilibrium equations associated with the Lagrangian $\E$ in \eqref{eq:combined_functional}.}

\subsection{Equilibrium equations}
The configuration space of an NPN ribbon, modulo translations and rigid rotations, is spanned by the functions $\{n_1,n_2,n_3,m_1,m_2,m_3,u_1,u_2,\eta,\mu,v_3\}$ where $n_i$ and $m_i$, with $i\in\{1,2,3\}$, are the director components of the internal force $\bn$ and internal moment $\bmo$. 
Their evolution in $s$ is governed by \eqref{eq:force_balance} and \eqref{eq:moment_balance} written in the director components, and subsequently transformed by introducing $\eta$,
\begin{subequations}\label{eq:balance_laws_director_frame}
	\begin{align}
		n_1' & = \eta u_1  n_2 - u_2 n_3 \, ,\label{eq:dn1}\\
		n_2' & = -\eta u_1 n_1 + u_1 n_3\, ,\label{eq:dn2}\\
		n_3' &= u_2 n_1 - u_1 n_2 \, ,\label{eq:dn3}\\
		m_1'&=\eta u_1 m_2 - u_2 m_3 + v_3 n_2\, ,\label{eq:dm1}\\
		m_2'&=-\eta u_1 m_1 + u_1 m_3 - n_1 v_3\, ,\label{eq:dm2}\\
		m_3'&=u_2 m_1 - u_1 m_2\, .\label{eq:dm3}
	\end{align}
\end{subequations}

Next we write down the constitutive equations for $m_i$ explicitly using \eqref{eq:combined_functional} and \eqref{eq:transformed_constitutive}.
We note that the square bracketed term in \eqref{eq:m1_transformed} can be eliminated using \eqref{eq:m3_transformed}.
From this, along with the relations $\partial\eta/\partial u_1 = -\eta/u_1$ and $\partial\eta/\partial u_3 = 1/u_1$, we obtain
\begin{align}
	m_1 + \eta m_3 = \frac{\partial\E}{\partial u_1}\, .\label{eq:m1_m3_relation}
\end{align}
Using \eqref{eq:combined_functional} in \eqref{eq:m1_m3_relation}, we arrive at the following explicit expression for $u_1$,
\begin{align}
	u_1 = \left(\frac{3+S_0}{3+S}\right)\frac{3v_3^2\left(m_1 + \eta m_3\right)}{8h^3\left(1+\eta^2\right)^2}\, .\label{eq:u1}
\end{align}
Similarly, by some calculations and use of \eqref{eq:combined_functional} and \eqref{eq:dm2}, 
equations \eqref{eq:m2_transformed} and \eqref{eq:m3_transformed} can be written as,
\begin{align}
	m_2 = \mu\, ,\label{eq:m2}
\end{align}
\begin{gather}
	m_3=\frac{1}{u_1}\left(\frac{3+S}{3+S_0}\right)\left\{\frac{16h^3\gamma u_1^2\eta(1+\eta^2)}{3v_3^2} + h\gamma \left[\left(\frac{1+S_0}{1+S}\right)\frac{2\eta}{v_3^2(1+\eta_0^2)} - \frac{2v_3^2\eta(1+\eta_0^2)}{(1+\eta^2)^2}\right.\right.\nonumber\\
	\left.\left. +\frac{2(1+\eta_0^2)}{v_3^2(1+S)}\left(\frac{\eta v_3^2}{1+\eta^2} - \frac{\eta_0}{1+\eta_0^2}\right) \left(v_3^2\frac{1-\eta^2}{1+\eta^2} + \frac{v_3^2\eta}{1+\eta^2} - \frac{\eta_0}{1+\eta_0^2}\right)\right]\right\}\nonumber\\
	+\frac{1}{u_1}\left\{2m_2\eta\left(\frac{u_2}{1+\eta^2} + \frac{v_3^2\eta_0'}{(1+\eta^2)^2}\right) - \frac{-\eta u_2 m_1 + u_1m_3 - n_1 v_3}{(1+\eta^2)}\right\}\,,\label{eq:m3}
\end{gather}
the latter of which makes also use of  the former.
Furthermore, the equilibrium equation corresponding to $\mu$ can be obtained from \eqref{eq:EL_function} as,
\begin{align}
	\eta' = u_2(1+\eta^2) + v_3^2\eta'_0\, ,\label{eq:deta}
\end{align}
which is simply the inextensibility constraint rearranged.
Finally, using \eqref{eq:EL_stretch} and \eqref{eq:combined_functional}, the equilibrium equation for the stretch $v_3$ can be written as,
\begin{gather}
	n_3 = \frac{3+S}{3+S_0}\bigg\{\frac{-8h^3 \gamma u_1^2(1+\eta^2)^2}{3v_3^3}+h\gamma\bigg[\frac{2v_3(1+\eta_0^2)}{1+\eta^2} + \left(\frac{1+S_0}{1+S}\right)\left(\frac{1+\eta^2}{1+\eta_0^2}\right)\left(\frac{-2}{v_3^2}\right)\nonumber\\
	+4\left(\frac{\eta v_3^2}{1+\eta^2} - \frac{\eta_0}{1+\eta_0^2}\right)\frac{\eta(1+\eta_0^2)}{v_3(1+S)} + \left(\frac{\eta v_3^2}{1+\eta^2} - \frac{\eta_0}{1+\eta_0^2}\right)^2\frac{(1+\eta^2)(1+\eta_0^2)}{1+S}\left(\frac{-2}{v_3^3}\right)\bigg]\bigg\}\nonumber\\
	+m_2\left(\frac{2v_3\eta'_0}{1+\eta^2}\right)\,.\label{eq:n3}
\end{gather}

\subsection{Kinematic equations}
The centerline of the ribbon can be obtained by integrating relation \eqref{eq:kinematics}$_1$.
To that end we parametrise the director frame attached to the centerline using the \emph{quaternion} parameter $q_i= q_i(s)$, where $i\in\{1,2,3,4\}$, subject to the constraint of unit norm $q_1^2+q_2^2+q_3^2+q_4^2=1$ (see, for example, \cite{dichmann:hamiltonian,champneys:elastic}).\footnote{Often the four rotation parameters $q_i$ are called the Euler-Rodrigues parameters in the literature. However, as noted in \cite{altmann:hamilton}, this attribution is disreputable.} 
The Cartesian components of the directors in a fixed lab frame $\{\be_1,\be_2,\be_3\}$ can then be represented as
\begin{gather}\label{eq:director_parametrisation}
	\bd_1=\begin{pmatrix}
		q_1^2-q_2^2-q_3^2+q_4^2\\
		2(q_1 q_2+ q_3q_4)\\
		2(q_1 q_3 - q_2 q_4)
	\end{pmatrix}\, ,\qquad
	\bd_2 =\begin{pmatrix}
		2(q_1 q_2 -  q_3 q_4)\\
		-q_1^2 + q_2^2 - q_3^2 + q_4^2\\
		2 (q_2 q_3 + q_1 q_4)
	\end{pmatrix}\, ,\nonumber\\
	\bd_3 = \begin{pmatrix}
		2(q_1 q_3 +  q_2 q_4)\\
		2(q_2 q_3 - q_1 q_4)\\
		-q_1^2 - q_2^2 + q_3^2 + q_4^2
	\end{pmatrix}\, ,
\end{gather}
Using the above in \eqref{eq:kinematics}$_2$, the evolution equations of $q_i$s can be obtained as,
\begin{align}\label{eq:euler_parameters_ODEs}
	\begin{pmatrix}
		q_1'\\
		q_2'\\
		q_3'\\
		q_4'
	\end{pmatrix} = \frac{1}{2}\begin{pmatrix}
		0 & u_3 & -u_2 & u_1\\
		-u_3 & 0 & u_1 & u_2\\
		u_2 & -u_1 & 0 &u_3\\
		-u_1 & -u_2 & -u_3 &0
	\end{pmatrix}
	\begin{pmatrix}
		q_1\\
		q_2\\
		q_3\\
		q_4
	\end{pmatrix}\, .
\end{align}
The Cartesian components of the centerline in a fixed lab frame, namely $\{r_1,r_2,r_3\}$ can then be obtained by integrating the following equations, which follow from  \eqref{eq:director_parametrisation}$_3$ and \eqref{eq:kinematics}$_1$,
\begin{subequations}
	\begin{align}\label{eq:centerline_ODEs}
		r'_1 &= 2v_3\left(q_1 q_3 + q_2 q_4\right)\,,\\
		r'_2 &= 2v_3\left(q_2 q_3 -  q_1 q_4\,,\right)\\
		r'_3 &=v_3\left(-q_1^2 - q_2^2 + q_3^2 + q_4^2\right)\,.
	\end{align}
\end{subequations}

\subsection{The full differential-algebraic system}
With the parameterisation of the directors $\bd_i$, the system of equations describing a thin NPN ribbon comprise 14 first order differential equations, and 2 algebraic equations.
The differential equations are the force and moment balance equations \eqref{eq:balance_laws_director_frame}, the incompressibility constraint \eqref{eq:deta}, and the kinematic equations  \eqref{eq:euler_parameters_ODEs} and \eqref{eq:centerline_ODEs},
whereas the algebraic equations are \eqref{eq:m3} and \eqref{eq:n3}.
For brevity of notation, we represent our system as,
\begin{subequations}\label{eq:dae_system}
	\begin{align}
		\bx' &= \bff(\bx,\by)\,,\label{eq:dae_system_1}\\
		{\bf 0}&=\bg(\bx,\by)\,,\label{eq:dae_system_2}
	\end{align}
\end{subequations}
where $\bx = \{r_1,r_2,r_3,q_1,q_2,q_3,q_4,n_1,n_2,n_3,m_1,m_2,m_3,\eta\}^T$ are the differential functions, whereas $\by=\{u_2,v_3\}^T$ are the algebraic variables.
The vectors $\bff$ and $\bg$ are respectively the right-hand side of equations \eqref{eq:balance_laws_director_frame}, \eqref{eq:deta}, \eqref{eq:centerline_ODEs}, \eqref{eq:euler_parameters_ODEs},  and \eqref{eq:m3} and \eqref{eq:n3}.

The system requires a total of 14 boundary conditions to be complete.
We postpone the specification of these conditions until Sect.~\ref{sec:a_cantilever_ribbon}.

The differential-algebraic system obtained thus far is of index 2 according to the definition in \cite{ascher:computer}, 
which means that the algebraic equations can be differentiated twice to obtain a set of ODE's for the algebraic functions $\by$.
Our attempt to obtain such a system led us to equations which contained expressions that were intractable and unmanageable.
To avoid such difficulties, we chose to  discretize directly the algebraic equations along with the differential ones using the method of orthogonal collocation on finite elements, the details of which are discussed in the next section.

\section{Numerical discretization}\label{sec:numerics}
We follow the lecture notes of Doedel \cite{doedel:lecture}
to discretize our system \eqref{eq:dae_system} using orthogonal collocation.
While Doedel's notes focus on system of ordinary differential equations only, we modify his approach to obtain a discretization for the accompanying algebraic equations as well.

We discretise the normalized domain $0\leq s\leq1$ into $N$ finite elements of equal size such that $0=s_0<s_1<s_2....s_N=1$,
and denote and define the $j$th element as $h_j:= s_j-s_{j-1}$, where $1\le j \le N$.
Our objective is to find  vectors of polynomials $\bp_h\in\mathbb{P}^m_h$ and $\bq_h\in\mathbb{P}^{m-1}_h$, where $\mathbb{P}^m_h$ is the space of vector piecewise polynomials such that the following collocation equations, 
\begin{align}\label{eq:collocation_equations}
	\bp'_h = \bff(\bp_h(z_{j,i}),\bq_h(z_{j,i})),\qquad {\bf 0}=\bg(\bp_h(z_{j,i})\,,\bq_h(z_{j,i}))
\end{align}
are satisfied with $\bp_h$ obeying the accompanying boundary conditions.
The collocation points $z_{j,i}$ in each sub-interval $h_j$ are the scaled roots of the $m$th degree element of a system of orthogonal polynomials (Gauss points).

In each sub-interval $[s_{j-1},s_j]$, the vector of polynomials are approximated as
\begin{align}
	\bp_j=\sum_{i=0}^m l_{j,i}(s)\bx_{j-\tfrac{i}{m}},\qquad\bq_j=\sum_{i=0}^{m-1} l_{j,i}(s)\by_{j-\tfrac{i}{m}}\,,
\end{align}
where $l_{i,j}$, with $j=1,\dots,N$ and $i=0,\dots,m$, are the Lagrange  polynomials defined as,
\begin{align}
	l_{i,j}=\prod_{k=0,k\ne i}^m \frac{s-s_{j-\frac{k}{m}}}{s_{j-\frac{i}{m}}-s_{j-\frac{k}{m}}}\, .
\end{align}
With this choice, we have
\begin{align}
	\bx_{j-\tfrac{i}{m}} = \bx\big(s_{j-\tfrac{i}{m}}\big),\qquad \by_{j-\tfrac{i}{m}} = \by\big(s_{j-\tfrac{i}{m}}\big)\,,
\end{align}
where $\bx(s)$ and $\by(s)$ are the solutions of the continuous problem.
Furthermore, we enforce the continuity of the polynomials $\bp_j$ across elements.
No such condition is imposed on the $\bq_j$'s approximating the algebraic functions.

The number of differential functions in our system is $14$ and the number of algebraic functions is $2$. 
Since a polynomial of degree $m$ is determined by $m+1$ parameters, the total number of degrees of freedom of all the polynomials approximating our system is $14(m+1)N + 2mN = 16mN+14N$.
This is equal to the number $14mN + 2mN$ of collocation equations \eqref{eq:collocation_equations}  plus the number $14(N-1)$ of continuity conditions for the differential functions  plus $14$, the number of boundary conditions on the differential variables.
Thus, our system is complete.

The resulting discretised algebraic equations are  solved by using a pseudo-arclength continuation scheme \cite{allgower:introduction}. 

\section{A cantilever ribbon}\label{sec:a_cantilever_ribbon}
We employ the apparatus developed so far to compute configurations of a rectangular ribbon of length $L$ and thickness $h$ under pure mechanical stretching and bending, and subsequent activation.
Our end objective is to compute estimates of the work that can be recovered from a mechanically deformed ribbon by activating it.
\begin{figure}[h!]
	\captionsetup[subfigure]{justification=centering}
	\centering
	\begin{subfigure}{0.48\textwidth}
		\centering
		\includegraphics[width=\textwidth]{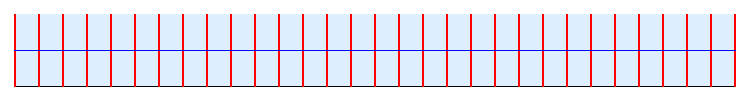}
		\caption{$n=0$}
		\label{fig:n0}
	\end{subfigure}
	\begin{subfigure}{0.48\textwidth}
		\centering
		\includegraphics[width=\textwidth]{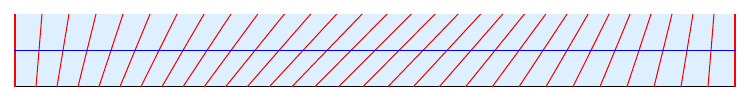}
		\caption{$n=1$}
		\label{fig:n1}
	\end{subfigure}
	
		\begin{subfigure}{0.48\textwidth}
		\centering
		\includegraphics[width=\textwidth]{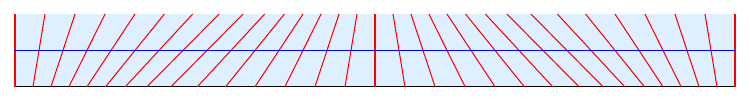}
		\caption{$n=2$}
		\label{fig:n2}
	\end{subfigure}
	\caption{ \alert{Three} distributions, governed by equation \eqref{eq:alpha_0}, of nematic directors imprinted on a narrow rectangular NPN ribbon. The width of the ribbon has been exaggerated for clarity of presentation.}
\end{figure}

We consider a ribbon for which the translational and rotational degrees of freedom at $s=0$ are constrained, and a dead load $\bP$ applied at the other end.
The boundary conditions to be imposed on system \eqref{eq:dae_system_1} are given by,
\begin{subequations}
	\begin{align}
		r_1(0)&=0\, ,\\
		r_2(0)&=0\, ,\\
		r_3(0)&=0\, ,\\
		q_1(0)&=0\, ,\\
		q_2(0)&=0\, ,\\
		q_3(0)&=0\, ,\\
		q_4(0)&=1\, ,\\
		n_1(0)&=\bP\cdot\be_1\, ,\\
		n_2(0)&=\bP\cdot\be_2\, ,\\
		n_3(0)&=\bP\cdot\be_3\, ,\\
		m_1(L)&=0\, ,\\
		m_2(L)&=0\, ,\\
		m_3(L)&=0\, ,\\
		\eta(0)&=0\, .
	\end{align}
\end{subequations}
The distribution of the nematic directors across the length are assumed such that the angle $\alpha_0(s)$ they make with $\be_3$ is given by
\begin{align}\label{eq:alpha_0}
	\alpha_0(s) = \frac{\pi}{4}\sin\left(\frac{n\pi s}{L}\right)\,,
\end{align}
where $n$ is an integer.
We will restrict our analysis to reference configurations with \alert{$n=0, 1,$ and $2$}.

For each distribution of the nematic directors, we conduct two numerical experiments, namely the \emph{pulling} experiment and the \emph{bending} experiment, the details of which are described in the following.

\subsection{The pulling experiment}
We consider an initially straight ribbon fixed at $s=0$ so that the director frame $\{\bd_1(0),\bd_2(0),\bd_3(0)\}$ coincides with the Cartesian basis $\{\be_1,\be_2,\be_3\}$, whose origin is placed at $\br(0)$.
The ribbon is mechanically stretched by a force $\bP=P\be_3$  applied at $s=1$, and gradually increased from $P=0$ to $P=6\times 10^{-3}\, kL^2$, where  $k$ is an elastic modulus of the ribbon selected so that $kL^3=E_0$, which by \eqref{eq:scaling_energy} amounts to set
\begin{equation}
	\label{eq:k_definition}
	k:=2\frac{w}{L}n_\mathrm{s}k_\mathrm{B}T_0.
\end{equation}
The order parameter $S$ is held constant at $S=S_0=4.788$ during this process.\footnote{This specific value of $S_0$ has been selected so as to correspond via \eqref{eq:Q_and_S} to a Maier-Saupe order parameter $Q_0\doteq0.61$, which minimizes $U_\mathrm{MS}$ for $\beta=7.5$; in explicit physical terms, this amounts to take the temperature $T_0\doteq0.91T_\mathrm{NI}$, where $T_\mathrm{NI}$ is the temperature of the nematic-to-isotropic transition of the nematogenic molecules (also known as the \emph{clearing} temperature).}
Subsequently, we activate the ribbon by varying \emph{quasi-statically}
the order parameter $S$ in the interval $[0,9]$, while holding the load $\bP$ constant.\footnote{In the variable $Q$, the interval $[0,9]$ for $S$ becomes the interval $[0,0.75]$, whose end-points are the absolute minimizers of $U_\mathrm{MS}$ for any $\beta\le6.81$ and $\beta\doteq9.05$, respectively.}
This is the general setting of the pioneering experiments reported in \cite{tajbakhsh:spontaneous} (see also \cite[p.\,69]{warner:liquid}).

\begin{figure}[t]
	\captionsetup[subfigure]{justification=centering}
	\centering
	\begin{subfigure}[t]{0.49\textwidth}
		\centering		\includegraphics[width=\textwidth]{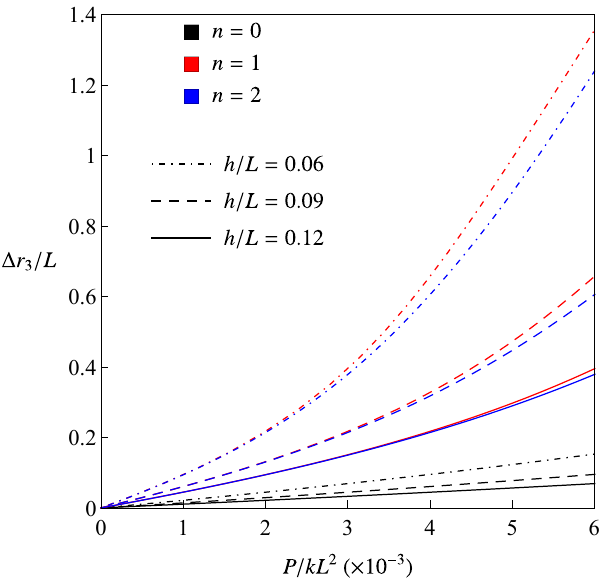}
		\caption{}\label{fig:pulling_experiment_a}
	\end{subfigure}
	\hfill
	\begin{subfigure}[t]{0.49\textwidth}
		\centering
		\includegraphics[width=\textwidth]{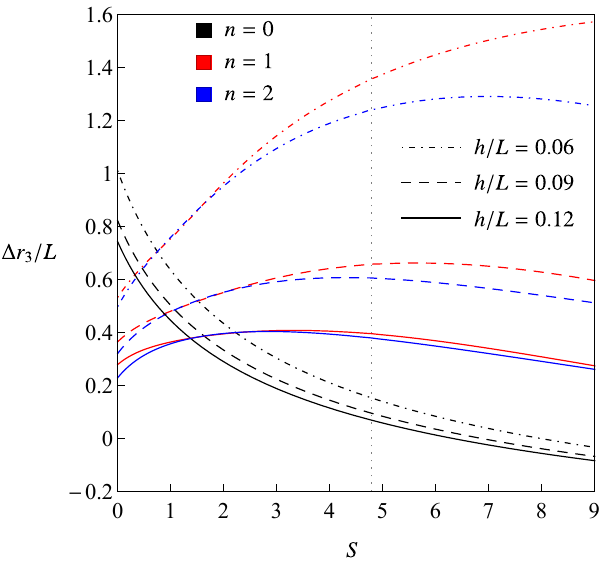}
		\caption{}
		\label{fig:pulling_experiment_b}
	\end{subfigure}     
	
	\begin{subfigure}[t]{0.3\textwidth}
		\centering
		\includegraphics[trim={1cm 0 2cm 0},clip,scale=0.4]{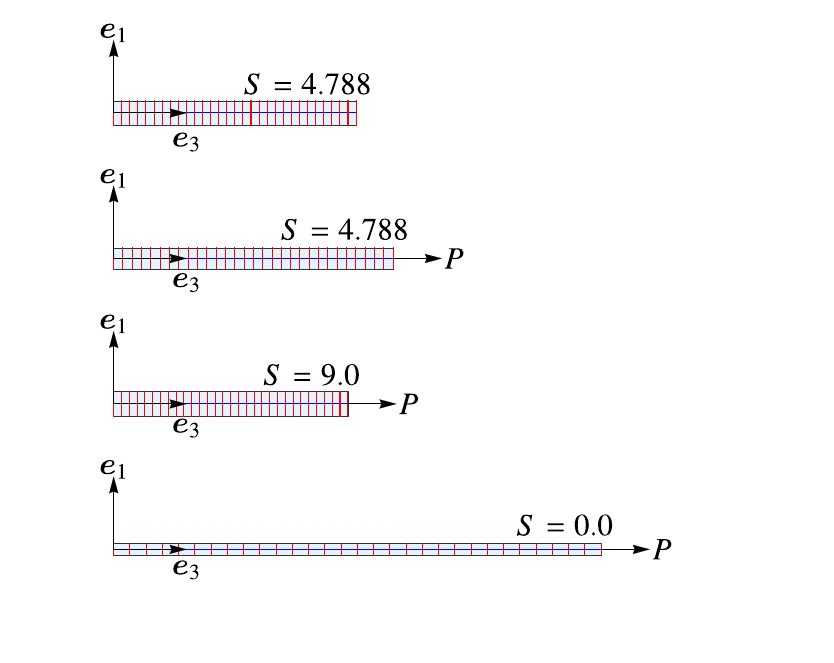}
		\caption{}
		\label{fig:pulling_experiment_c}
	\end{subfigure} 
	\begin{subfigure}[t]{0.3\textwidth}
		\centering
		\includegraphics[trim={1cm 0 0.4cm 0},clip,scale=0.4]{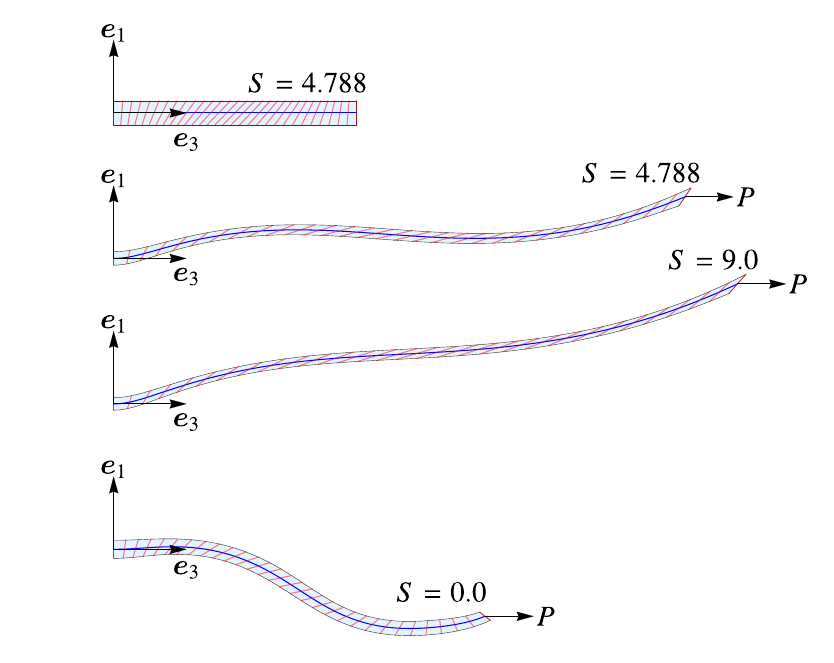}
		\caption{}
		\label{fig:pulling_experiment_d}
	\end{subfigure}    
	\hspace{1cm}
	\begin{subfigure}[t]{0.3\textwidth}
		\centering
		\includegraphics[trim={1cm 0 1cm 0},clip,scale=0.4]{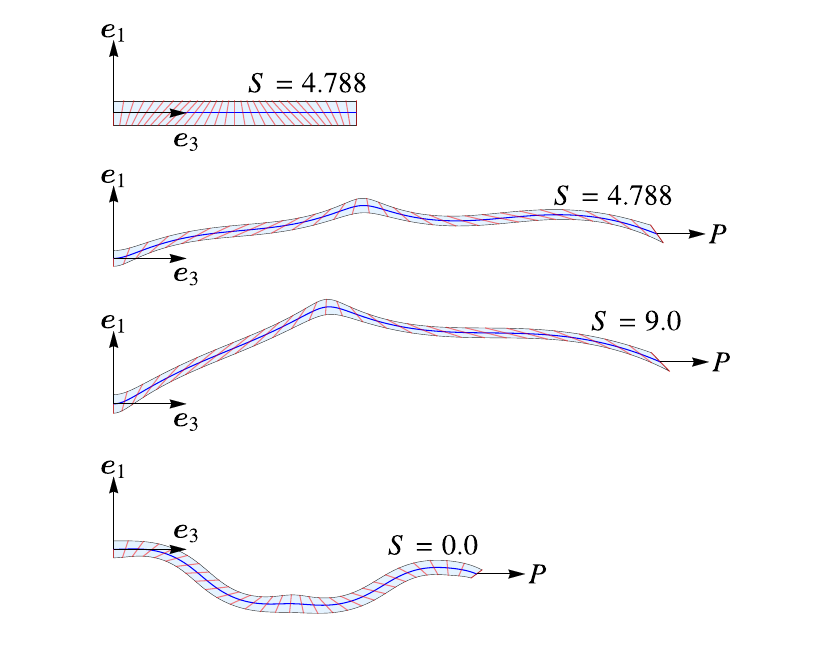}
		\caption{}
		\label{fig:pulling_experiment_e}
	\end{subfigure}   
	\caption{The figure pertains to various results from the pulling experiment with $P=6\times 10^{-3}kL^2$. \eqref{fig:pulling_experiment_a} Plots of the normalized displacement $\Delta r_3/L$ of the tip of the NPN ribbon as a function of the applied force $P/kL^2$ for various values of the thickness $h/L$. \eqref{fig:pulling_experiment_b} Plots of $\Delta r_3/L$ as a function of the activation parameter $S$ (with $S_0=4.788$), obtained by activating the configurations which have been pulled. Panels \alert{\eqref{fig:pulling_experiment_c}, \eqref{fig:pulling_experiment_d} and \eqref{fig:pulling_experiment_e}}  show deformed configurations of the ribbon of thickness $h/L=0.06$ for the cases \alert{$n=0$, $n=1$ and $n=2$}, respectively.}
	\label{fig:pulling_experiment}
\end{figure}

Fig.~\ref{fig:pulling_experiment_a} shows the plot of the net normalized tip displacement of the ribbon ($\Delta r_3/L$) against the normalized\footnote{\alert{Scaling the force as $P/kLh$ would result in the collapse of all force-displacement curves for a given distribution $n$, since in the pulling experiment the energy scales like $h$. The resulting plot is shown in the  Fig.~\ref{fig:pulling_experiment_f} of Appendix~\ref{sec:pull_LD_scaled}. A similar collapse, however, would not take place for the bending experiment. To treat both experiments on equal footing, we have retained for both the scaling of the load that we deem most natural.}} applied force $P/kL^2$ for distributions of the nematic director field corresponding to $n=0$, $n=1$, \alert{$n=2$}, and different values of the normalized thickness $h/L$.
Since a higher thickness leads to an effectively higher axial stiffness of the ribbon, thicker ribbons undergo smaller tip displacements as compared to thinner ribbons for \alert{ $n=0$,  $n=1$, and $n=2$}.
For a given load, the net displacement of the tip remains consistently higher for \alert{both $n=1$ and $n=2$} as compared to $n=0$, indicating that a varied distribution of the nematic directors along the length reduces the axial stiffness of the ribbon.
It is worth pointing out that since the case $n=0$ possesses symmetry about the plane spanned by $\{\be_2,\be_3\}$, the centerline remains straight upon deformation (see Fig.~\ref{fig:pulling_experiment_c}).
This symmetry is broken for \alert{both $n=1$ and $n=2$}, resulting in a deformation of the centerline in the $\{\be_1,\be_3\}$ plane upon pulling (Fig.~\ref{fig:pulling_experiment_d}).
\alert{The deformation  of the centerline in the $\{\be_1,\be_3\}$ plane for $n=2$ results in concentration of in-plane curvature in certain regions, while no such concentration is noted for $n=1$.
The tip displacement for $n=1$ is observed to be higher than for the case $n=2$, indicating that stronger variations in the nematic directors tend to result in stiffer axial response.} 
We also observe that the reduction in the axial stiffness for \alert{both $n=1$ and $n=2$,} as compared to $n=0$ becomes more pronounced as the load increases. 

Next we activate the deformed ribbon, with $P=6\times 10^{-3}\, kL^2$ applied to it, from its base state $S_0$ by varying $S$ from $S_0$ to $0$ and from $S_0$ to $9$.
The change in the tip displacement $\Delta r_3/L$ as a function of $S$ is shown in Fig.~\ref{fig:pulling_experiment_b}.
Here we observe a qualitative difference between the response of the ribbon for $n=0$ \alert{in comparison to  $n=1$ and $n=2$.}
For $n=0$, where the nematic directors are aligned orthogonally to the centerline in the reference configuration, the values of $S>S_0$ lead to a retraction of the ribbon tip in the direction opposite to the applied force. 
For $S<S_0$, we observe the opposite behavior, i.e., the ribbon further elongates under the same load upon activation.
This observation can be understood in light of the fact that an increase in $S$ over $S_0$ means an increase in the nematic order, which is accompanied by a dilation of the fibers along the nematic directors.
Since the deformation is assumed to be area preserving, this leads to the shortening of the fibers orthogonal to the nematic directors, which in this case happens to be along the centerline.

For \alert{both $n=1$ and $n=2$}, the behavior of the ribbon is qualitatively different from the case with $n=0$.
The tip of the ribbon consistently retracts against the applied load for $S<S_0$ for all the three thicknesses studied.	
For $S>S_0$, whether the tip displaces along or against the load depends on the thickness $h/L$ of the ribbon.
We observe that for $h/L=0.06$, the tip displaces along the load $\bP$ for $S>S_0$, whereas for $h/L=0.09$ and $0.12$, there is a net retraction of the tip against the load.
We believe that this behavior is a result of the complex interplay between the dilation/contraction of the fibers, and the shear deformation that they undergo due to the variation of the nematic director field along the length of the ribbon. 
\alert{Here again, the $n=2$ ribbons have  more pronounced  deformation of the center-line in the $\{\be_1,\be_3\}$ plane and  smaller tip displacements as compared to $n=1$ }.
\alert{Figs.~\ref{fig:pulling_experiment_c},~\ref{fig:pulling_experiment_d} and~\ref{fig:pulling_experiment_e}} show various configurations of the ribbon corresponding to mechanical stretching and subsequent activation for \alert{$n=0$, $n=1$ and $n=2$}, respectively.

\subsection{The bending experiment}
Next we explore the behavior of the NPN ribbon to bending out of plane.
We apply a transverse point load $\bP=P\be_2$ at $s=1$, and vary its magnitude from $P=0$ to $P=6\times 10^{-4}\, kL^2$, while the other end remains fixed as in the pulling experiment.
The order parameter $S$ is held constant at $S_0$ during this process.
\begin{figure}[h!]
	\captionsetup[subfigure]{justification=centering}
	\begin{subfigure}{0.473\textwidth}
		\centering
		\includegraphics[width=\textwidth]{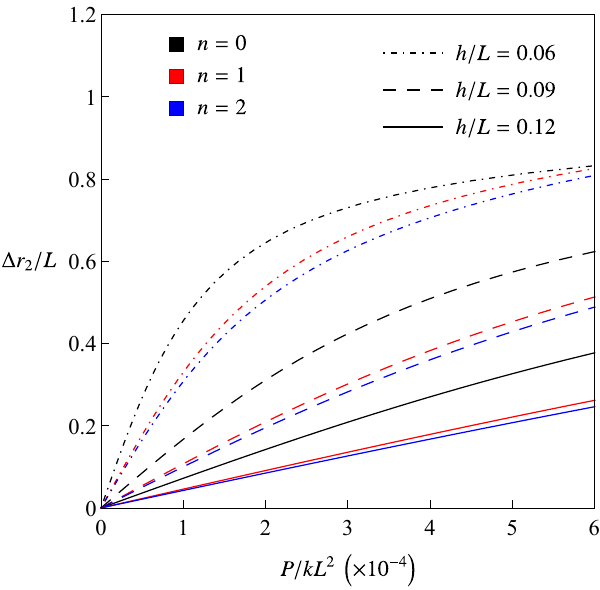}
		\caption{}
		\label{fig:bending_experiment_a}
	\end{subfigure}
	\hfill
	\begin{subfigure}{0.475\textwidth}
		\centering
		\includegraphics[width=\textwidth]{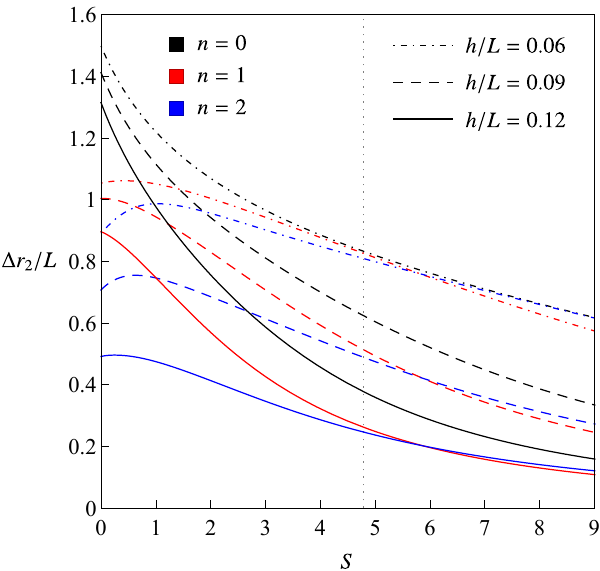}
		\caption{}
		\label{fig:bending_experiment_b}
	\end{subfigure}     
	
	\hspace{-1cm}
	\begin{subfigure}{0.3\textwidth}
		\centering
		\includegraphics[scale=0.3]{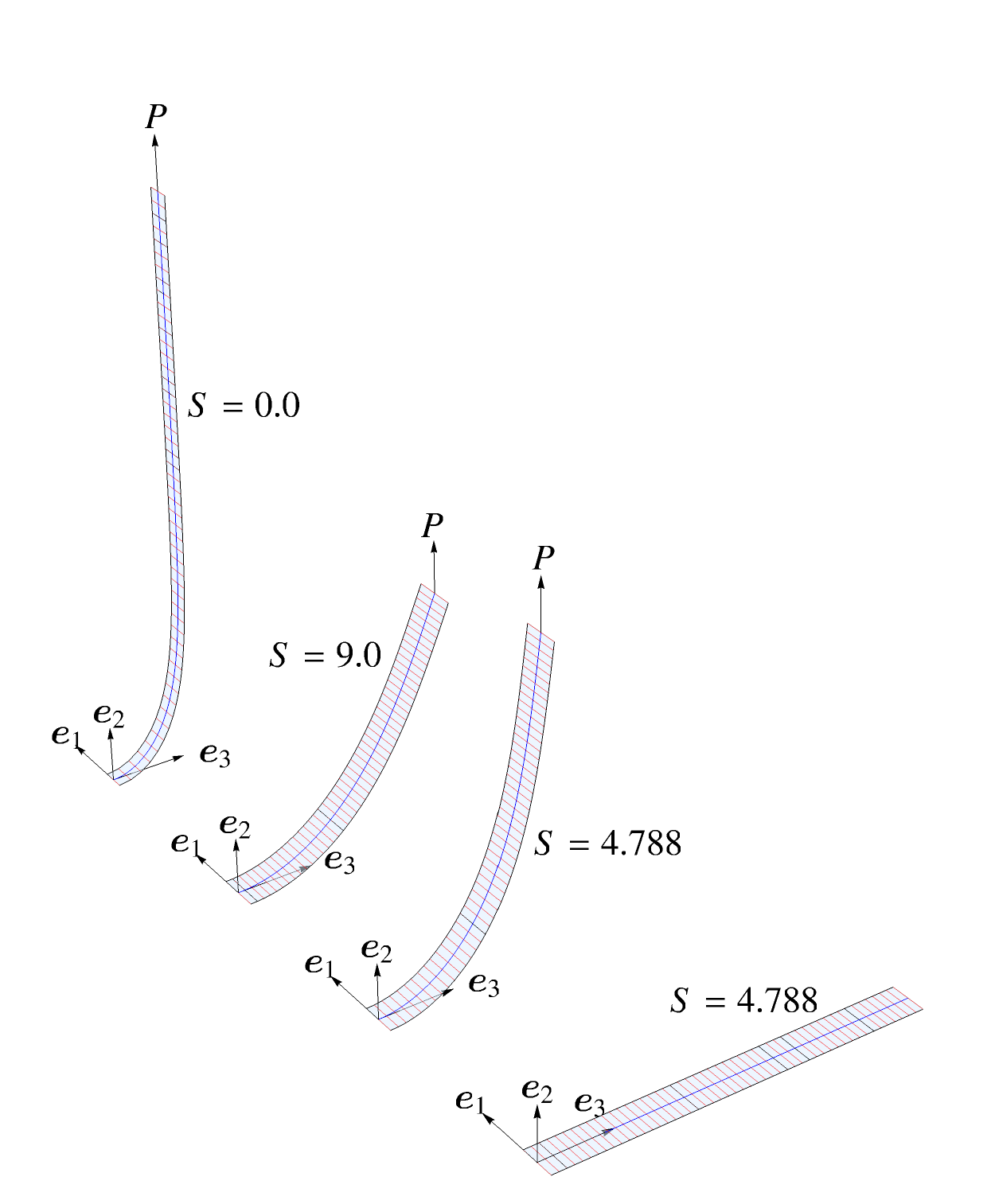}
		\caption{}
		\label{fig:bending_experiment_c}
	\end{subfigure}  
	\hspace{0.4cm}   
	\begin{subfigure}{0.3\textwidth}
		\centering
		\includegraphics[scale=0.3]{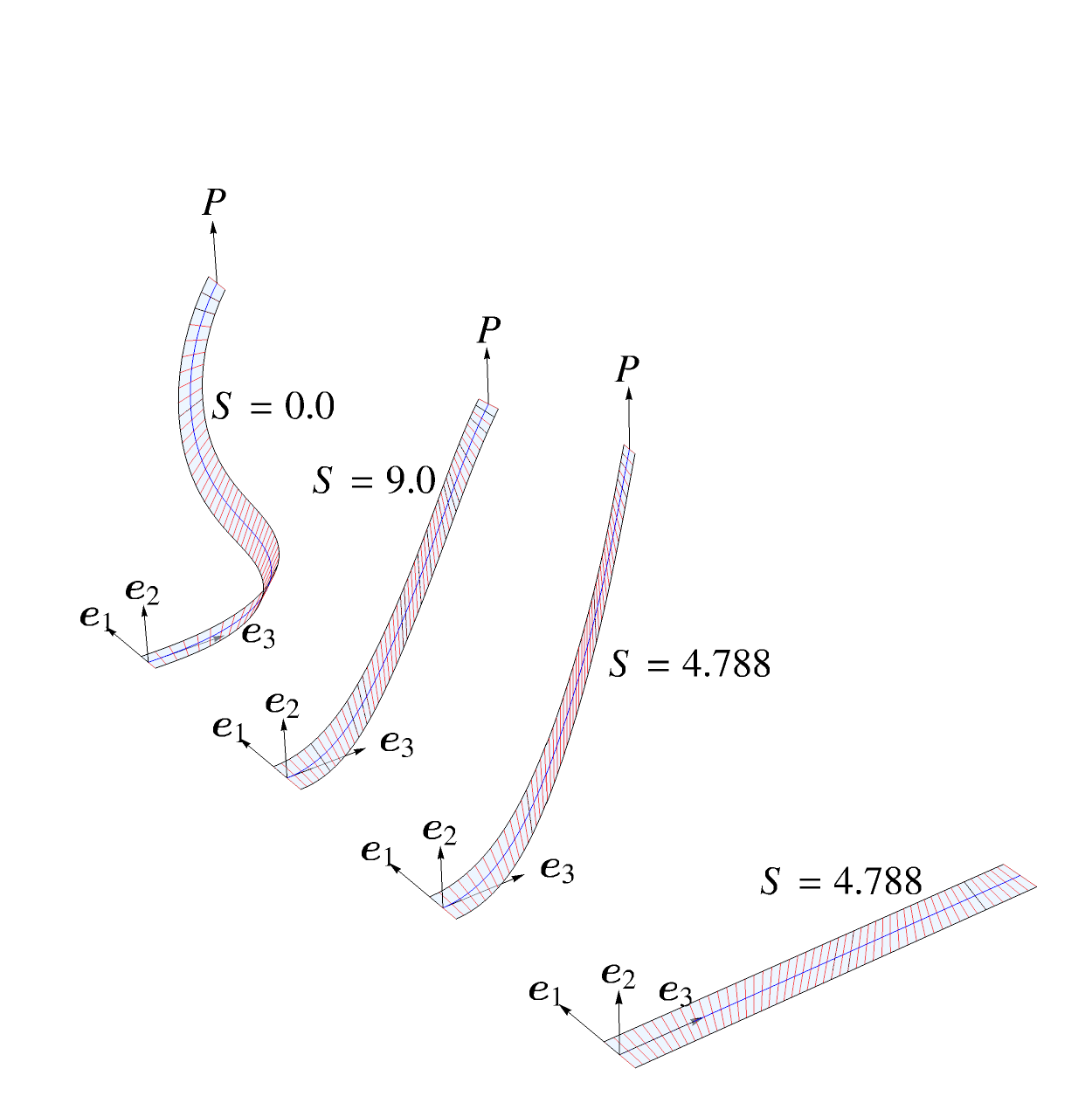}
		\caption{}
		\label{fig:bending_experiment_d}
	\end{subfigure}    
	\hspace{0.7cm}
		\begin{subfigure}{0.3\textwidth}
		\centering
		\includegraphics[scale=0.3]{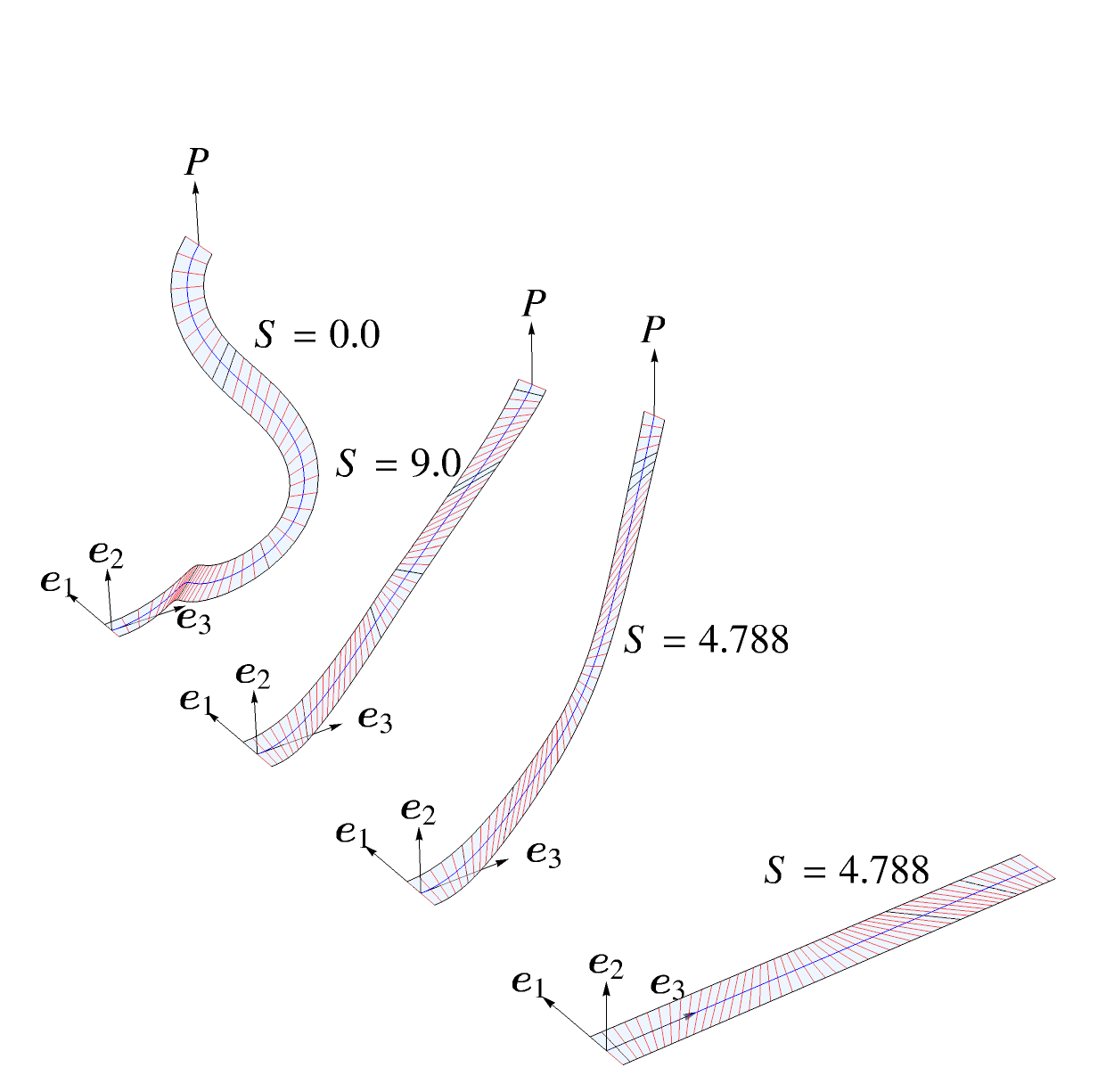}
		\caption{}
		\label{fig:bending_experiment_e}
	\end{subfigure}    
	\caption{\eqref{fig:bending_experiment_a} Plots of the normalized displacement $\Delta r_2/L$ of the tip against the applied transverse load $P=6\times 10^{-4} kL^2$ for various values of the thickness $h/L$. \eqref{fig:bending_experiment_b} The transverse displacement $\Delta r_2/L$ plotted against the activation parameter $S$ (with $S_0 = 4.788$) obtained upon activating the configurations under mechanical bending. \alert{Panels \eqref{fig:bending_experiment_c}, \eqref{fig:bending_experiment_d} and \ref{fig:bending_experiment_e}} show the deformed configurations of the ribbon of thickness $h/L=0.06$ in the bending experiment for the cases \alert{$n=0$, $n=1$ and $n=2$}, respectively.} 
\end{figure}
The normalized component of the transverse displacement $\Delta r_2/L$ of the tip  is plotted against the normalized load $P/kL^2$ in Fig.~\ref{fig:bending_experiment_a}.
Contrary to the pulling experiment, we observe that the transverse displacement for the \alert{ cases $n=1$ and $n=2$} are consistently lower than for $n=0$ for all thicknesses.
This indicates that the anisotropy in this case produces an increase in  the effective bending stiffness of the ribbon.
It is also worth noting that for $n=0$ the ribbon strictly bends in the $\{\be_2,\be_3\}$ plane, whereas for \alert{ $n=1$ and $n=2$ the ribbon also twists in addition to bending (see Figs.~\ref{fig:bending_experiment_c}, \ref{fig:bending_experiment_d} and \ref{fig:bending_experiment_e}).}

The deformed ribbon with a transverse tip load of $P=6\times 10^{-4}kL^2$ is then activated by varying the order parameter $S$ from its base state $S_0$ in the range $S\in[0,9]$.
Fig.~\ref{fig:bending_experiment_b} shows the variation of the normalized tip displacement as a function of the order parameter $S$.
As opposed to the pulling experiment, the response of the ribbon to activation remains qualitatively the same for \alert{$n=0$, $n=1$ and $n=2$}.
In  \alert{all three cases}, the ribbon retracts for $S>S_0$, whereas it deflects further for $S<S_0$.

\section{Activation efficiency}\label{sec:efficiency}
We showed in the preceding section that a terminally loaded NPN ribbon can pull back on axially and transversely applied terminal point loads, if activated the right way.
The activation of an NPN ribbon is achieved by driving the scalar order parameter $S$ away from its value $S_0$ in the reference configuration to a target value chosen so as to recover mechanical work.
\alert{For the pulling experiment, the chosen target values of $S$ are, respectively, $S=9$ for  for $n=0$  and $S=0$ for $n=1$ and $n=2$.}
For the bending experiment, the chosen target value for $S$ is $S=9$ for \alert{all} nematic director arrangements.
Such changes in the order  parameter requires energy exchange between the ribbon and its environment in form of heat or light.
While our theory is agnostic to the mode  of  energy exchange, we assume, for definiteness, that the activation  is brought about by an exchange of heat.
We now compute the amount of heat required to drive $S$ from $S_0$ to a target value.
We quantify the capacity of an NPN ribbon to yield workby defining the \emph{activation} efficiency $\mathsf{e}$ as,
\begin{equation}
	\label{eq:tm_efficiency}
	\mathsf{e}:=\left| \frac{W_A}{\Delta Q_\mathrm{h}}\right|\,,
\end{equation}
where $W_A<0$ is the mechanical yield of activation, that is, the  work done by the ribbon against the point force during activation, and $\Delta Q_\mathrm{h}$ is the amount of heat injected into the ribbon to activate it.
The mechanical yield of activation is evaluated as,
\begin{align}
	W_A &= \bP\cdot\int_{\br_P(L)}^{\br_A(L)} d\br=\bP\cdot\left(\br_A(L)-\br_P(L) \right) \, ,\label{eq:work_activation}
\end{align}
where $\br_P(L)$, and $\br_A(L)$ are the positions of the free end of the ribbon centerline in the pulled, and fully activated configurations, respectively.
The work is computed only for those situations where the activation leads the tip of the ribbon to displace against the applied load, and is therefore expected to be negative.
For instance, in the pulling experiment, $W_A$ in \eqref{eq:work_activation} is computed for $S$ going from $4.788$ to $9$ for $n=0$, whereas for $n=1$ \alert{ and $n=2$} it is computed for $S$ going from $4.788$ to $0$.
For the bending experiment, $W_A$ is computed for $S$ going from $4.788$ to $9$ for \alert{ all three cases $n=0$, $n=1$ and $n=2$}.

We compute $\Delta Q_\mathrm{h}$ for a given activation process from the following statement of balance of energy,
\begin{align}
	E_{A} = E_{P} + W_A+\Delta Q_\mathrm{h}\, , \label{eq:total_energy}
\end{align}
where $E_{A}$ and $E_{P}$ are the total energies of the ribbon (obtained by integrating \eqref{eq:combined_functional} over the centerline) in the activated state and the purely mechanically pulled/bent state respectively.

\begin{figure}[h]
	\captionsetup[subfigure]{justification=centering}
	\begin{subfigure}{0.45\textwidth}
		\centering
		\includegraphics[width=\textwidth]{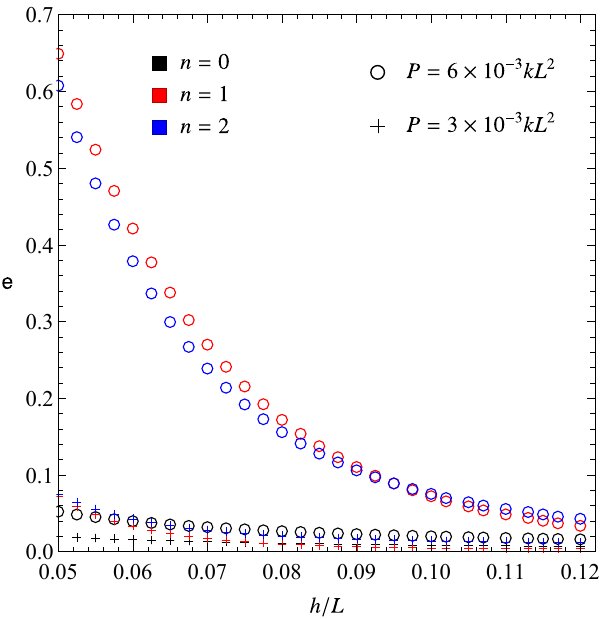}
		\caption{}
		\label{fig:efficiency_a}
	\end{subfigure}
	\hfill
	\begin{subfigure}{0.45\textwidth}
		\centering
		\includegraphics[width=\textwidth]{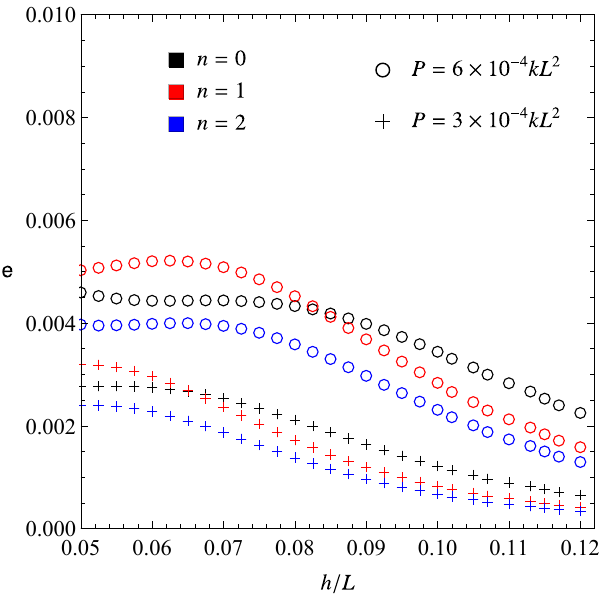}
		\caption{}
		\label{fig:efficiency_b}
	\end{subfigure}     
	\begin{subfigure}{0.45\textwidth}
		\centering
		\includegraphics[width=\textwidth]{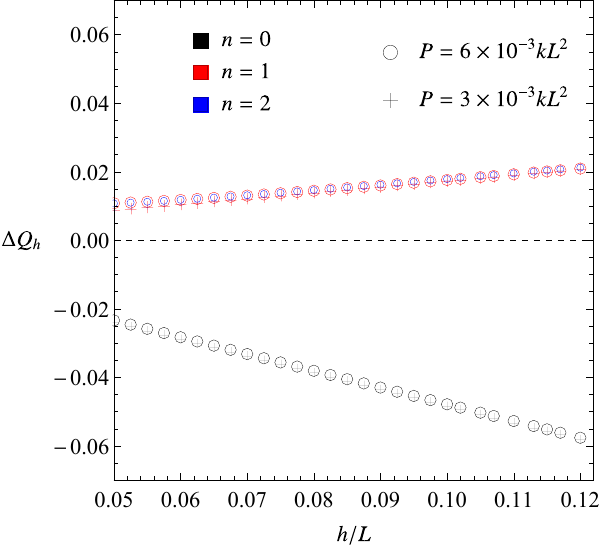}
		\caption{}
		\label{fig:heat_c}
	\end{subfigure}
	\hfill
	\begin{subfigure}{0.45\textwidth}
		\centering
		\includegraphics[width=\textwidth]{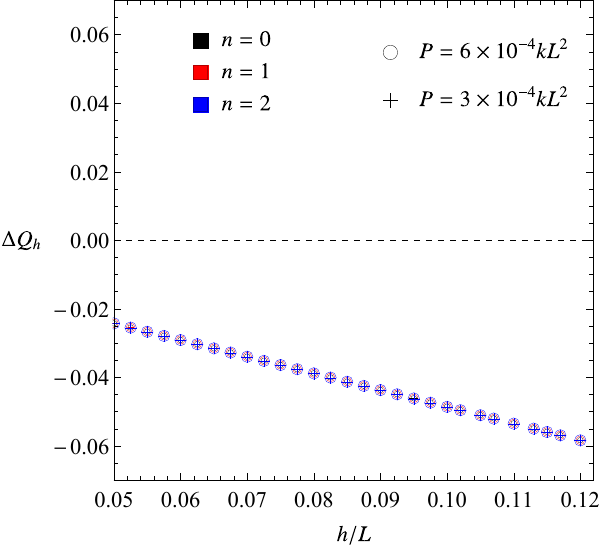}
		\caption{}
		\label{fig:heat_d}
	\end{subfigure}     
	\caption{Panels \ref{fig:efficiency_a} and \ref{fig:efficiency_b} show how the activation efficiency $\mathsf{e}$  of an NPN ribbon defined in \eqref{eq:tm_efficiency} varies with the normalised thickness ($h/L$) for the pulling and bending experiment, respectively. Panels \ref{fig:heat_c} and \ref{fig:heat_d} show the amount of heat $\Delta Q_\mathrm{h}$ (scaled to $2kL^2 w$) as a function of the normalised thickness  of the ribbon ($h/L$) in pulling and bending, respectively.}
	\label{fig:energy_computations}
\end{figure}

We present the  efficiency $\mathsf{e}$ defined in \eqref{eq:tm_efficiency} for both the pulling and the bending experiments against the (scaled) thickness $h/L$ of the ribbon in Figs.~\ref{fig:efficiency_a} and \ref{fig:efficiency_b}, respectively.
In addition, we note the change in efficiency in response to two values of the applied loads$, P=6\times 10^{-3} kL^2$ and $P=3\times 10^{-3} kL^2$ for both pulling and bending experiments.

\alert{In the pulling experiment with the higher load the efficiencies observed are consistently higher than those for the lower load for all thicknesses.
For the lower load, while the efficiencies for $n=2$ are  consistently higher than for $n=0$ and $n=1$, the two efficiency curves for $n=0$ and $n=1$, instead, cross  at a critical thickness of approximately $h/L\approx 0.0725$, above which efficiency is higher for $n=0$ than for $n=1$.  
We also note that for the higher load, efficiencies for $n=1$ \alert{and $n=2$} are consistently higher than for $n=0$. 
At the same time, a crossing in the efficiency curves for $n=1$ and $n=2$  is observed at a critical thickness of  $h/L\approx 0.0925$, below which efficiency for  $n=1$ is higher than for $n=2$.
For the bending experiment, we observe that higher values of the dead load result in higher efficiencies for both $n=0$ and $n=1$. 
However, for each value of the dead load, the same crossing in the efficiency curves seen for the lower load in the pulling experiment manifests itself: there are two critical thicknesses, $h/L\approx0.065$ and $h/L\approx0.0825$ (for the lower and higher loads, respectively), above which the efficiency is again higher for $n=0$ than for $n=1$. The efficiency for $n=2$ is consistently lower than for $n=0$ and $n=1$ for both loads.
}

\alert{Minor details apart, the efficiency curves exhibit a major difference for pulling and bending experiments. In the former, a higher efficiency under a \emph{longitudinal} load is associated with the presence of  distortion \emph{along} the ribbon. In the latter, the efficiency is practically insensitive to longitudinal distortion. An increase in bending efficiency under a \emph{transverse} load could likewise be expected if the nematic distortion were increased \emph{across} the ribbon's thickness, but our theory is presently incapable of capturing this feature.
	
In general, we may say that an increase in the applied load increases systematically the efficiency  of thermal recovery under load, as does an increase of distortion along the applied load.} 

We also note that an increase in the value of $S$ from its value $S_0$ in the reference configuration requires heat to be extracted from the ribbon, whereas a decrease in $S$ requires heat to be injected.
Figure~\ref{fig:heat_c} shows $\Delta Q_\mathrm{h}$ as a function of the normalized thickness in the activation process of  the ribbon, starting from its mechanically fully pulled configuration. 
Computation of $\Delta Q_\mathrm{h}$ from \eqref{eq:total_energy}
for the pulling experiment shows that for $n=1$ \alert{and $n=2$}, heat must be injected into the ribbon to recover work (i.e., $\Delta Q_\mathrm{h}>0$), whereas for $n=0$, heat must be extracted (i.e., $\Delta Q_\mathrm{h}<0$).
This observation is consistent with the fact that the order parameter $S$ had to be increased from its reference value for $n=0$ in order to recover work, whereas for $n=1$ \alert{and $n=2$} it had to be decreased.

Figure~\ref{fig:heat_d} shows $\Delta Q_\mathrm{h}$ as a function of the normalized thickness of the ribbon for the bending experiment. In this case,
for \alert{all three $n=0$, $n=1$ and $n=2$}, heat must be extracted (i.e. $\Delta Q_\mathrm{h} <0$) from the ribbon to activate it in a fashion that the resulting deformation pulls back on the applied load.
This is consistent with Fig.~\ref{fig:bending_experiment_b}, where in the bending experiment the order parameter $S$ is increased from its base value  $S_0$ in order to recover work.
We also note that the numerical value of $\Delta Q_\mathrm{h}$ happens to be very close for $n=0$, $n=1$ \alert{and $n=2$}. 

Figures~\ref{fig:heat_c} and \ref{fig:heat_d} reveal that $\Delta Q_\mathrm{h}$ is practically unaffected by the load $P$ and it scales linearly with the thickness $h$ (that is, with the volume of the three-dimensional ribbon), in complete agreement with our (approximate) representation of the activation process a one traversing a sequence of equilibrium condensation states identified with the absolute minimizers of $U_\mathrm{MS}$ induced by compliant choices of $\beta$ in \eqref{eq:U_MS}. This way of reasoning is further illuminated in Appendix~\ref{sec:toy}, where it is applied to a simplified setting.

\section{Conclusions}\label{sec:conclusions}
We computed the equilibrium configurations of a ribbon consisting of nematic polymer networks under the action of an external load. The numerical code that was instrumental to this end allowed us to depict the swirling configurations that the ribbon can achieve, most of them exhibiting various degrees of  bending and twisting.

Our major interest lied in describing the mechanical behaviour of a ribbon activated under the persistent action of a load. In particular, we performed two numerical experiments:  in one, the load pulled the ribbon; in the other, it bent it. In both experiments, we imprinted \alert{three} different nematic textures in the reference configuration (which were conveyed by the ensuing deformations): one was uniform, the other was not. The activation of the material was devised in such a way that the spontaneous deformation induced by it would antagonize the applied load, so as to recover part of the work done by the latter.  We defined the efficiency of the activation process as the ratio between the work recovered and the activation energy.

\alert{Our elementary experiments suggest two general conjectures about the activation efficiency of pre-loaded NPN ribbons, one more general than the other: (1) efficiency increases with the load, (2) non-uniformity in the imprinted nematic texture increases efficiency when it unfolds along the load.}

Further studies are needed to either prove or disprove these conjectures in more general settings, as well as to evolve into dynamics the customary quasi-static approach to actuation adopted here.

\backmatter

\bmhead{Acknowledgments}
\alert{We are grateful to an anonymous Reviewer, whose comments and suggestions have prompted an improvement of our work. E.G.V. is a member of \emph{GNFM}, a division of  \emph{INdAM}, the Italian Institute for Advanced Mathematics.}

\section*{Declarations}

\subsection*{Competing interests}
The authors declare to have no competing interests.

\begin{appendices}
\section{Alternate scaling } \label{sec:pull_LD_scaled}
\begin{figure}[H]
	\centering
	\includegraphics[scale=0.6]{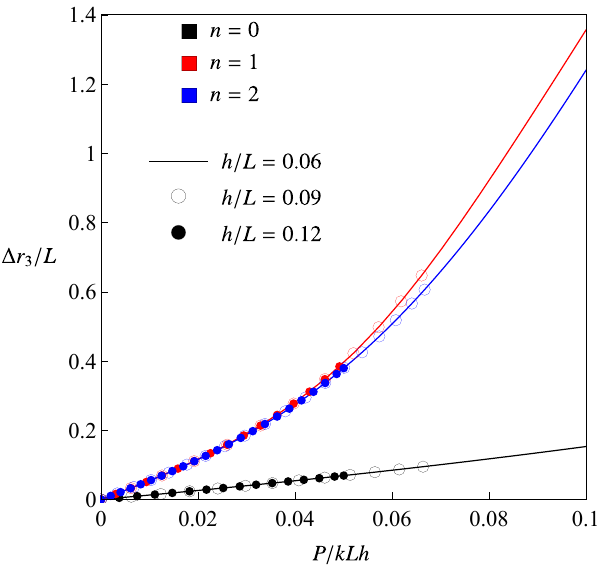}
	\caption{Plots of the normalized displacement $\Delta r_3/L$ of the tip of the NPN ribbon as a function of the applied force $P/kLh$ for various values of the thickness $h/L$.}
	\label{fig:pulling_experiment_f}
\end{figure}

\alert{In Fig.~\ref{fig:pulling_experiment_f}, we plot the force-displacement curves for the pulling experiment by scaling the applied load $P$ to $kLh$ instead of $kL^2$ (as done in Fig.~\ref{fig:pulling_experiment_a}): for a given nematic distortion, all curves collapse on the same curve, irrespective of the ribbon's thickness, as, in the absence of bending, the whole elastic energy scales like $h$.}

\section{Toy model}\label{sec:toy}
Consider a system with a single mechanical variable, $x$, and a single internal variable, $S$. The internal energy of the system has two components, $E(x,S)$, which depends on both variables, and $U(S,S_0)$, which depends on $S$ and the value $S_0$ of $S$ that makes $U$ attain its absolute minimum, so that
\begin{equation}
	\label{eq:toy_internal_energy_minimum}
	\frac{\partial U}{\partial S}\bigg|_{S=S_0}=0,\quad \frac{\partial^2 U}{\partial S^2}\bigg|_{S=S_0}>0\,.
\end{equation}
Moreover, the power expended by an external (dead) load $P$ is given by $P\dot{x}$, where a superimposed dot denotes differentiation with respect to time.

Equilibrium requires the function $L:=E+U-Px$ to be stationary, that is,
\begin{equation}
	\label{eq:toy_equilibrium}
	\frac{\partial E}{\partial x}=P,\quad\frac{\partial E}{\partial S}+\frac{\partial U}{\partial S}=0\,.
\end{equation}
We denote by $(x_\mathrm{eq},S_\mathrm{eq})$ the pair of solutions to these equations corresponding to the absolute minimum of $L$, assumed to exist. Both $x_\mathrm{eq}$ and $S_\mathrm{eq}$ will be regarded as functions of $S_0$.

Now, we imagine a quasi-static process that takes the system through a set of equilibrium states parameterized by a curve $t\mapsto S_0(t)$. The first law of thermodynamics dictates that during this process the following equation be obeyed,
\begin{equation}
	\label{eq:toy_first_law}
	(E+U)^{\dot{\null}}_{(x_\mathrm{eq},S_\mathrm{eq})}=P\dot{x}_\mathrm{eq}+Q_\mathrm{h},
\end{equation}
where $Q_\mathrm{h}$ is the energy supplied to the system per unit time (possibly in form of heat) to perform the process. In \eqref{eq:toy_first_law}, both $x_\mathrm{eq}$ and $S_\mathrm{eq}$ depend on $t$ through $S_0$. It follows from \eqref{eq:toy_first_law} that 
\begin{equation}
	\label{eq:toy_first_law_consequence}
	\frac{\partial}{\partial t}[E(x_\mathrm{eq},S_\mathrm{eq})+U(x_\mathrm{eq},S_\mathrm{eq})-Px_\mathrm{eq}]=Q_\mathrm{h}\,.
\end{equation}
Making use of \eqref{eq:toy_equilibrium} in \eqref{eq:toy_first_law_consequence}, we readily obtain that
\begin{equation}
	\label{eq:toy_heat_power}
	\bigg(\frac{\partial U}{\partial S_0}\bigg)_{(x_\mathrm{eq},S_\mathrm{eq})}\dot{S}_0=Q_\mathrm{h}\,,
\end{equation}
and integrating  over the whole process, we finally arrive at
\begin{equation}
	\label{eq:toy_total_heat}
	\Delta Q_\mathrm{h}=\int\bigg(\frac{\partial U}{\partial S_0}\bigg)_{(x_\mathrm{eq},S_\mathrm{eq})}dS_0,
\end{equation}
which delivers the total amount of energy that needs to be supplied to the system during the imagined process.

Equation \eqref{eq:toy_total_heat} simplifies even further if we can assume that 
\begin{equation}
	\label{eq:toy_approximation}
	S_\mathrm{eq}\approx S_0\,,
\end{equation}
which is exact when $E$ does not depend on $S_0$. When \eqref{eq:toy_approximation} applies, the left-hand side of \eqref{eq:toy_heat_power} is approximately a total derivative and \eqref{eq:toy_total_heat} becomes
\begin{equation}
	\label{eq:toy_total_heat_approximation}
	\Delta Q_\mathrm{h}\approx\Delta U_\mathrm{min}\,,
\end{equation}
where $\Delta U_\mathrm{min}$ is the increment (with sign) of the minimal value of $U$ during the whole process. When applied to a three-dimensional body, \eqref{eq:toy_total_heat_approximation} implies that $\Delta Q_\mathrm{h}$ scales like the volume of the body, a conclusion confirmed by panels \ref{fig:heat_c} and \ref{fig:heat_d} of Fig.~\ref{fig:energy_computations} in the main text.

As an easy application of \eqref{eq:toy_total_heat}, we consider the case of a linear spring with rest length $\bar{x}$, whose preferred value $x_0$ minimizes an internal potential $U$. Here $\bar{x}$ and $x_0$ play the role of $S$ and $S_0$, respectively. We set
\begin{equation}
	\label{eq:toy_application}
	E=\frac12k(x-\bar{x})^2,\quad U=\frac12A(\bar{x}-x_0)^2 +B(x_0)\,,
\end{equation}
where $k>0$ is an elastic constant, $A>0$ is a material constant and $B$ is an increasing function. Simple calculations show that for this choice of functions \eqref{eq:toy_total_heat} can be cast in the following equivalent form,
\begin{equation}
	\Delta Q_\mathrm{h}=\Delta B-P\Delta x_0\,.
\end{equation}
For example, for $B=hx_0$ with $h>0$, \eqref{eq:toy_application} shows that for a given amount of heat $\Delta Q_\mathrm{h}$ poured into a pulled spring ($P>0$), the extension $\Delta x_0$ of the rest length produced by a process of  quasistatic thermal expansion is larger than for a compressed spring ($P<0$).

\end{appendices}


\end{document}